\documentclass[twocolumn,trackchanges]{aastex631}
\newcommand\ngc{NGC~1275}
\newcommand\fermi{{\it Fermi}}
\newcommand\gr{$\gamma$-ray}

\shorttitle{A Double-period QPO in NGC 1275}
\shortauthors{Zhang et al.}
\graphicspath{{./}{figures/}}

\begin{document}

\title{A double-period oscillation signal in millimeter emission of the radio galaxy NGC 1275}

\author{Pengfei Zhang}
\affiliation{Department of Astronomy, School of Physics and Astronomy, Key Laboratory of Astroparticle Physics of Yunnan Province, Yunnan University, Kunming 650091, People’s Republic of China; zhangpengfei@ynu.edu.cn; wangzx20@ynu.edu.cn}

\author[0000-0003-1984-3852]{Zhongxiang Wang}
\affiliation{Department of Astronomy, School of Physics and Astronomy, Key Laboratory of Astroparticle Physics of Yunnan Province, Yunnan University, Kunming 650091, People’s Republic of China; zhangpengfei@ynu.edu.cn; wangzx20@ynu.edu.cn}
\affiliation{Shanghai Astronomical Observatory, Chinese Academy of Sciences, 80 Nandan Road, Shanghai 200030, China}

\author{Mark Gurwell}
\affiliation{Center for Astrophysics $\mid$ Harvard \& Smithsonian,
60 Garden Street, Cambridge, MA 02138, USA}

\author{Paul J. Wiita}
\affiliation{Department of Physics, The College of New Jersey, 2000 Pennington Road, Ewing, NJ 08628-0718, USA}





\begin{abstract}

The nearby Seyfert type galaxy NGC~1275 contains a bright radio nucleus at 
its center, revealed through high-spatial resolution imaging to be the source 
of the jets emanating from the galaxy.  Coincident with the emergence of a 
new component C3 in the nucleus since 2005, flux densities from \ngc, at least 
at radio, millimeter (mm), and \gr\ frequencies, had been increasing up through 
2017 and leveled off afterwards. We analyze the long-term light curves 
	of the nucleus that span the rising trend to 2015 July, 
	and find a pair of 
approximately year-long quasi-periodic 
	oscillations, with periods of $P_l\simeq 345$\,d and
	$P_h\simeq 386$\,d respectively, in emission at 1.3-mm wavelength. 
	We discuss the case that there would be a long precession period 
	$P_{\rm prec}\simeq 9$\,yr, causing the appearance of $P_h$
	that is slightly higher than $P_l$. The
	accretion disk around the central supermassive black hole (SMBH) would
	be precessing at $P_{\rm prec}$, induced by either the Lense-Thirring
	effect or the existence of a companion SMBH. In the two
	scenarios, $P_l$ would be the jet wobbling timescale or
	the SMBH binary period respectively.
The finding, which could be verified through high-spatial resolution mm 
	imaging, would not only identify the nature of the jet variation
	but also help reveal the full features of the galaxy.

\end{abstract}


\section{Introduction}
\label{sec:intro}

The galaxy \ngc, with its radio counterpart named 3C~84, is located at 
the center of the Perseus cluster. As a cluster's central galaxy that is
close-by ($z = 0.0176$), it has been extensively observed in various studies. 
The galaxy itself has been classified as a Seyfert 1.5 type based on the
emission lines present in its optical spectrum \citep{vv10}. In its central
active nucleus, the supermassive black hole (SMBH)  
has a mass of $\sim 3.4\times 10^8$\,$M_{\odot}$, as
estimated from kinematic measurements of the surrounding molecular gas 
obtained from near-infrared observations \citep{wej05}. 
The nucleus contains jets seen at radio frequencies along the
south and north directions \citep{vrb94}. The radio core 
has been further resolved to mainly consist of two components, C1 and C3,
over the past 15 years (see Figure~\ref{fig:rimg}). 
While the C1 region is considered to be the base of the jet close to 
the center of the galaxy, the C3 component 
emerged in 2005 and has been moving towards the south
 \citep{nag+10}. In
addition, high-energy and very high-energy \gr\, emission has also been 
detected from NGC~1275 \citep{abd+09,ale+12}, presumably arising from the jets.

\begin{figure}
\centering
\epsscale{1.1}
\plotone{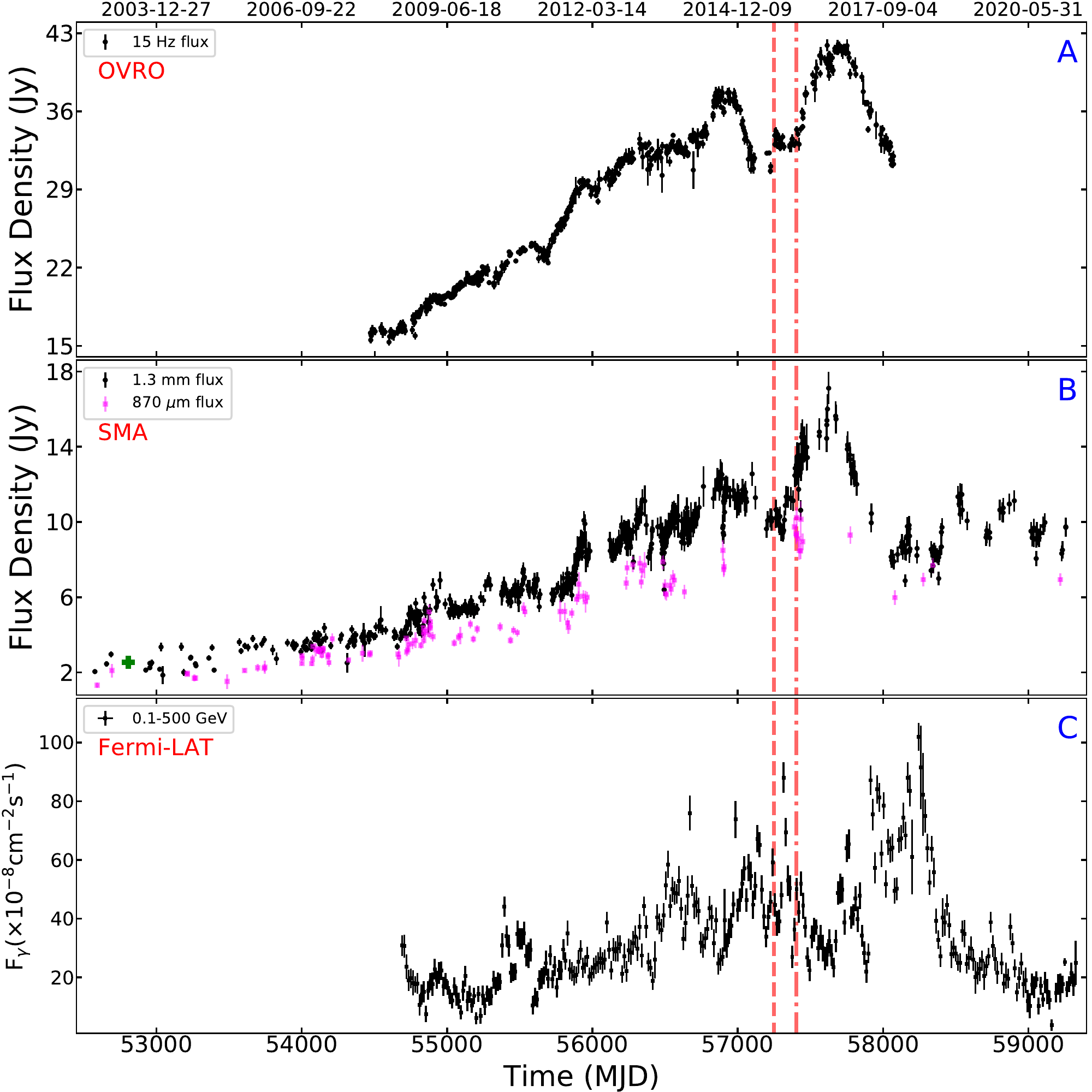}
	\caption{OVRO 15\,GHz ({\it panel A}), SMA 1.3-mm and 870-$\mu$m
	(black and pink respectively; {\it panel B}), and {\it Fermi}-LAT
	0.1--500\,GeV ({\it panel C}) light curves. The maximum interval
	of two adjacent data points in the 1.3-mm light curve is 239\,d, 
	marked by a green cross
	in panel B. The dashed and dash-dotted lines indicate 
	the times of 2015 Aug.\ 1 and 2016 Jan.\ 15. respectively 
	(cf.\ Section~\ref{sec:psd}). The radio and mm flux density peaks
	in 2016--2017 have been attributed to a collision between the 
	C3 jet and a clump of gas \citep{kin+18}.
\label{fig:lcs}}
\end{figure}

The core appeared to have had a rising trend in brightness 
at multiple wavelengths in 2003--2017 (\citealt{hod+18,bri+19,gul+21}; 
see  Figure~\ref{fig:lcs}).
This long-term
trend, as well as flare-like variations of \ngc, were revealed by
high-resolution imaging to primarily
arise from the emergence and flux density changes of the C3 
component \citep{nag+10,hod+18,kin+18}.
However according to the studies at 2--14\,mm reported by
\citet{hod+18} for the time period of 2013--2017, the C1 component could be
brighter and more variable at 2--3\,mm than C3. Also during 2015 August--September,
a ``flip'' in the direction of C3 was seen, followed by a monotonic flux density 
increase \citep{kin+18}. The flux density peaks in 2016--2017 seen in 
the long-term radio and mm light curves (cf., Figure~\ref{fig:lcs}) likely were
the result of the flip, thought
to be caused by the collision between the the tip of jet at C3 and a dense clump \citep{kin+18}.

Related to these intriguing results for the core of the NGC~1275 
jets, hints of precession of the jets have been reported.  
The derived precession timescales, however, have a very wide range: 
from as long as $\sim 33$\,Myr to explain the observed multi-kpc scale radio
(and X-ray) bubbles \citep{dfs06}; to $\sim 40$--100\,yr
from parsec-scale morphological changes and long-term flux density
variations of the jets \citep{bri+19}; $\sim 28.8$\,yr, from determining
the projected position angle of the C1 component \citep{dom+21}; 
or $\sim$6\,yr,
from a possible helical path of C3 (with respect to C1; \citealt{hiu+18}). 

Since the multiwavelength emission from the core of the jets in \ngc\ has 
been bright, monitoring of it has made long-term light curve data available 
at several wavelengths.
We took advantage of these data, and conducted searches for possible 
quasi-periodic oscillation (QPO) signals for a further study of the jets.
Surprisingly, a double-period oscillation signal at 1.3 mm wavelength was
found.  Previously, a search
for QPO signals was conducted with the 9-yr \gr\ data obtained
using the {\it Fermi Gamma-ray Space Telescope (Fermi)}, but no periodicities 
were found \citep{ndp20}.  

In the following Section~\ref{sec:dr}, we describe the data and periodicity
search analysis and provide the periodicity results. In Section~\ref{sec:dis},
the origin of the periodicities is discussed. Additional information about
the source and  results of our analyzes are provided in Appendices 
~\ref{sec:a1} and \ref{sec:a2}.

\section{Data analysis and Results}
\label{sec:dr}
\subsection{Data description and {\it Fermi} \gr\ data analysis}
The archival data we used to search for periodicity signals include 1.3-mm and
870-$\mu$m monitoring of \ngc\ with the Submillimeter Array 
(SMA; \citealt{gur+07}), 
15\,GHz monitoring with the Owens Valley Radio Observatory (OVRO; 
\citealt{rea+89,ric+11}), 
and 0.1--500\,GeV monitoring with the Large Area Telescope 
(LAT; \citealt{atw+09}) onboard \fermi.

The SMA has been monitoring over 400 sources on a several-day
cadence at these two wavelengths since 2002, and
\ngc\ (3C~84) is one of them. The two flux density light curves obtained
span over 18 yr, approximately from 2002 Nov. to 2021 Jan./Feb..
For the 1.3-mm light curve (Figure~\ref{fig:lc}), the median and mean 
time intervals between two
adjacent data points are 3.13 and 9.35\,d respectively, and the maximum
interval is $\sim$239\,d around MJD~52808 (2003 June 18). On many days two or
three measurements were made. For the 870-$\mu$m light curve 
(Figure~\ref{fig:lcs}),
the sampling cadence is much sparser compared to that of the 1.3-mm one: the
median, mean, and maximum time intervals between two adjacent data points
are 15.47, 60.24, and 876.65\,d, respectively.

The OVRO 40-m telescope has been monitoring over 1800 AGN
on a $\sim$3\,d cadence since 2008. \ngc\ (J0319+4130
in the OVRO target list) is one of the targets. The flux density light curve
(Figure~\ref{fig:lcs})
we used is approximately 10\,yr long, from 2008 Jan.\ 8 (MJD~54473) to 2017 
Nov.\ 15 (MJD~58072).
The median and mean intervals between two adjacent data points are 3.71 and
5.36\,d, respectively, and the maximum interval is 84.75\,d.
\begin{figure}
\begin{center}
\includegraphics[width=1.\linewidth]{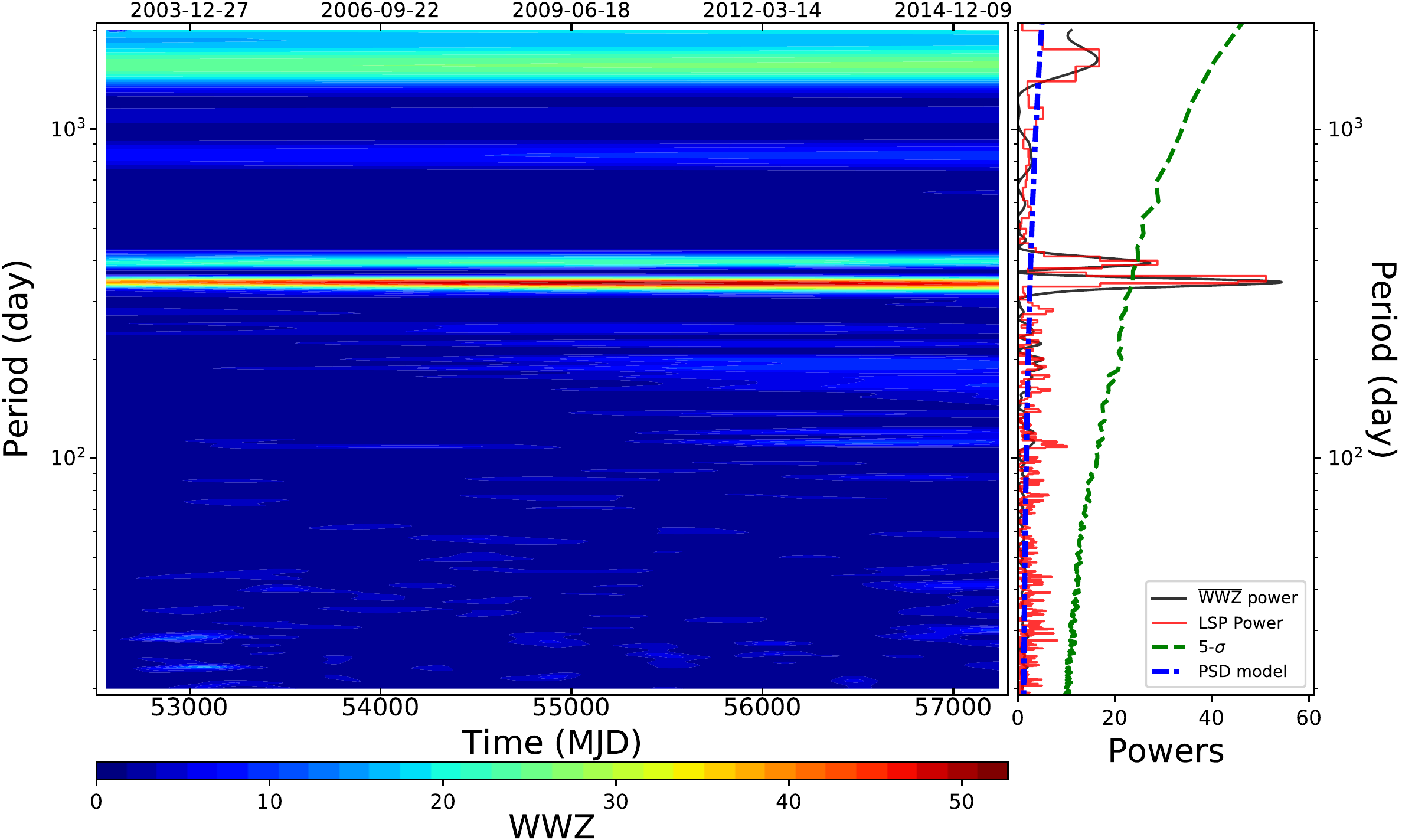}

\includegraphics[width=0.9\linewidth]{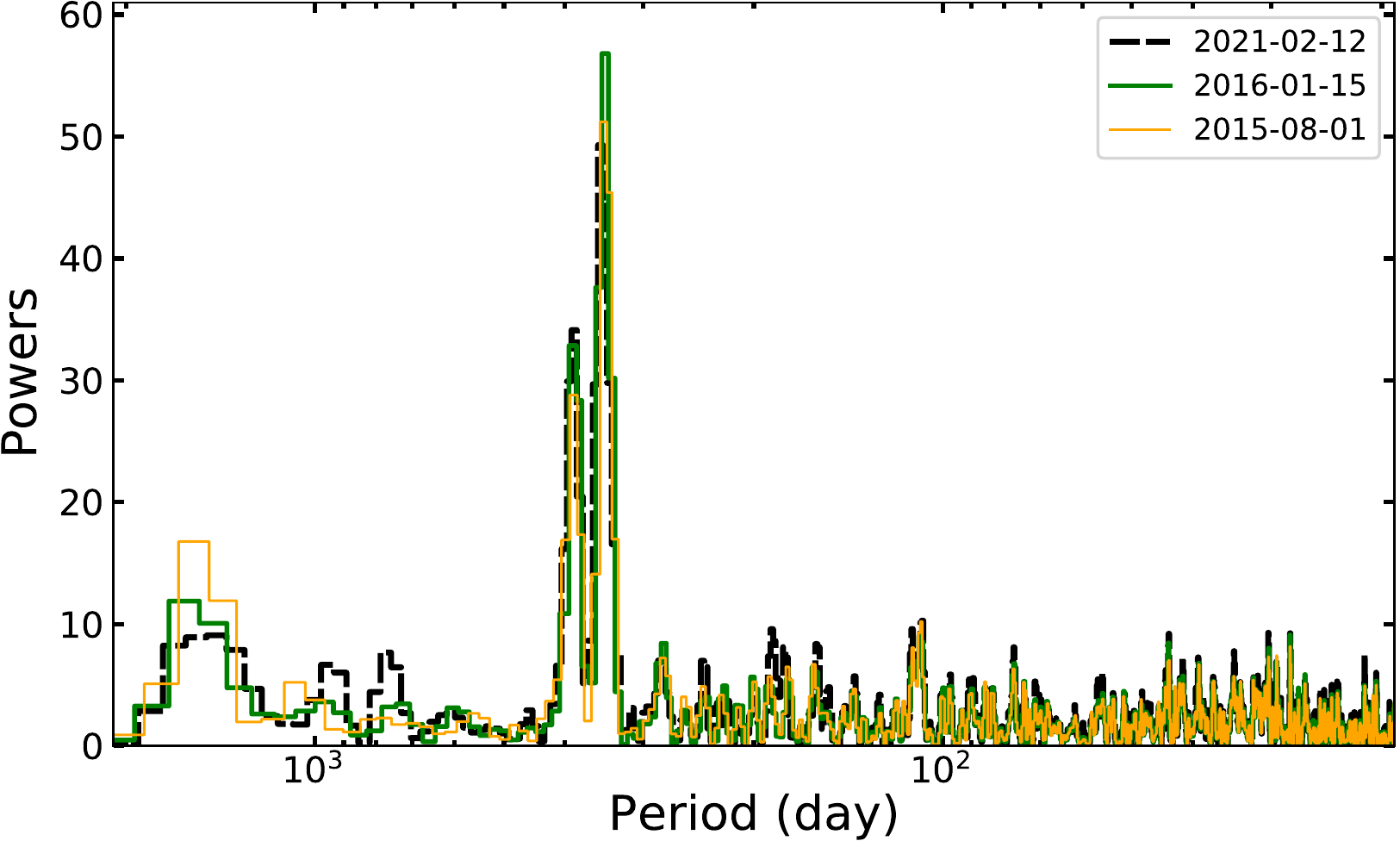}
\end{center}
	\caption{ {\it Upper left panel:} WWZ power for the 1.3-mm light 
curve in 2002--2015-July. Two periodicities, around $\sim 345$\,d and
$\sim 386$\,d, are clearly seen.
	{\it Upper right panel:} LSP power (red line) and summed WWZ power (black line) 
obtained for the 1.3-mm
light curve. The best fit to the underlying power spectral density is shown
as the blue dash-dotted line.
The 5$\sigma$ significance curve, obtained from the light curve simulation,
is shown as the green dashed curve. The power peaks for the two periodicities
are above the 5$\sigma$ significance curve. After taking into account 
a trial number
	of 350, the 345\,d (386\,d) periodicity still has a significance 
	greater 
	than (close to) 5$\sigma$.
	{\it Bottom panel:} Comparison of LSP powers from 
	the 1.3-mm light curve in 
	the time periods from 2002 to 2021 (black dashed), 
	2002 to 2016 Jan.\ 15 (green), 
	and 2002 to 2015 Aug.\ 01 (yellow). These power peaks are nearly the same, though 
	the highest one for the 346\,d periodicity is from the 2002 to 2016 
	Jan.\ 15 time period.
\label{fig:pow}}
\end{figure}

For the \emph{Fermi}-LAT data, we selected the 0.1-500~GeV Pass 8 
\emph{Front+Back} SOURCE class photon-like events from 2008 Aug.\ 4 
to 2021 Apr.\ 16
in a region of $20^{\circ}\times20^{\circ}$ centered at the position of \ngc.
In the fourth \emph{Fermi} Large Area Telescope catalog 
(4FGL; \citealt{abd+20}),
the source is named 4FGL~J0319.8+4130 with the coordinate
R.A.=$03^h19^m49\fs8$, Decl.=$41^{\circ}30^{'}43\farcs57$.
The events were reduced by selecting the zenith angle
$\rm <90^\circ,~DATA\_QUAL > 0, and ~LAT\_CONFIG=1$ to obtain high-quality
data in the good time intervals.
We built a model file that contained the spectral parameters of
all known 4FGL sources in the selected region.
The spectral form of \ngc\ in the 4FGL is log-parabola, described by
the form
$dN/dE = K (E/E_0)^{-\alpha_s -\beta_s \rm{log}(E/E_b)}$.
The binned maximum likelihood analysis was first applied to the whole
data, and for \ngc, we obtained an average photon flux
of $\rm (341.7\pm3.1)\times10^{-9}~photons\,cm^{-2}\,s^{-1}$ in the
0.1--500~GeV band with an overall test statistic (TS) value of $\sim$147500.
The corresponding best-fit spectral parameters we obtained 
are $\alpha_s = 2.078\pm0.005$,
 $\beta_s = 0.054\pm0.003$, and $E_b = 0.953\pm0.010$\,GeV,
in good agreement with that reported in 4FGL.

The results from the above analysis were saved as a new model file, based on
which the maximum likelihood analysis was performed again for
constructing a 0.1--500\,GeV light curve of \ngc\ 
(Figure~\ref{fig:lcs}).
We chose 15\,d as the time-bin size, so that the TS values obtained from
each time-bin data are mostly $\geq 25$. 

\subsection{Periodicity search and determination}
\label{sec:psd}

Two methods, the Weighted Wavelet Z-transform (WWZ; \citealt{fos96}) and 
the generalized Lomb-Scargle Periodogram (LSP; \citealt{lom76,sca82,zk09}) 
were employed to
search for periodicities in the radio, mm, and \gr\ light curves. Only
in the 1.3-mm light curve did we detect a significant QPO signal, which
actually consists of two periodicities (upper panels of Figure~\ref{fig:pow}).
This QPO signal is present throughout the entire light curve
(i.e., from 2002 to 2021). However, since the broad flux density peaks in
2016--2017 at radio and mm bands were likely due to the jet-clump collision
in 2015 Aug.--Sept., we also analyzed only the 2002--2015 Aug.\ 01 light curve data, and
the resulting power peaks are nearly the same as those from the whole
data (bottom panel of Figure~\ref{fig:pow}). In addition, we found that
the power peak at the lower period value actually strongest
when the data restricted to the slightly longer period of 2002--2016 Jan.\ 15 
are used. In any case, because of the evidence for a ``flip''  seen in high-resolution
imaging \citep{kin+18},  in the remainder of this paper we focus on the results from the 2002--2015 July
data. 

The period values of the two periodicities were obtained
by fitting the two LSP power peaks with Gaussian functions. The values are
$P_l = 344.6\pm10.0$\,d and $P_h = 386.0\pm10.9$\,d, 
where the uncertainties were taken as the half-widths at 
half-maximum of each peak.

In other light curves,
weak signals were seen in the 870-$\mu$m light curve, but
they did not have significances sufficiently high to make any QPO claims (see 
Appendix~\ref{sec:a2} and Figure~\ref{fig:pows}). We note that 
the 870-$\mu$m light curve only has 111 data
points, and is much sparser than the 1.3-mm light curve, so the lack of
a significant signal is not surprising.
\begin{figure}
\centering
\includegraphics[width=0.95\linewidth]{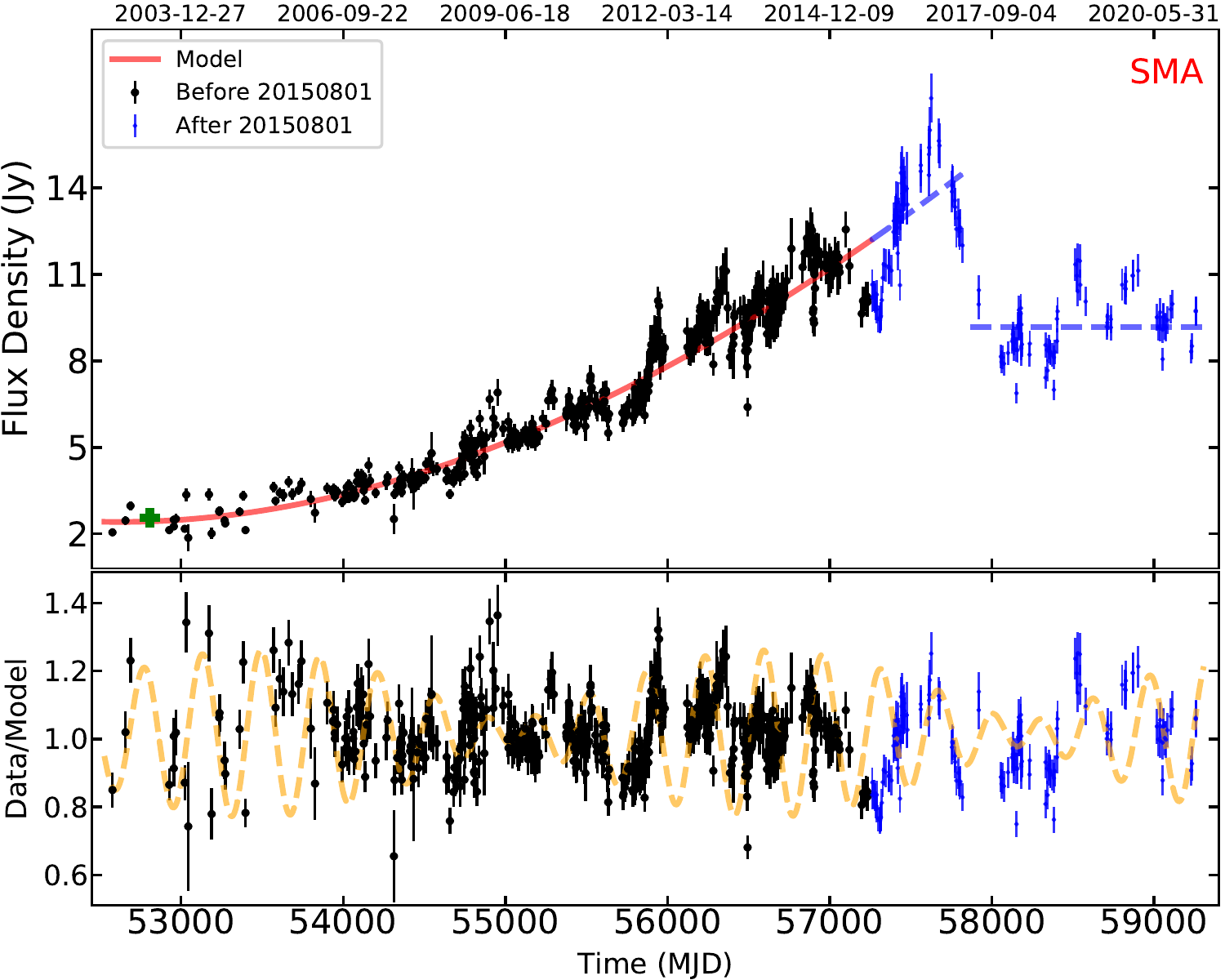}
\caption{{\it Top panel:} 
Long-term 1.3-mm light curve of \ngc. The fits given in 
	Section~\ref{sec:psd} were used to detrend the long-term
	variations in the light curve, which are a degree 3 polynomial
	(red and blue dashed lines) and a constant (horizontal blue dashed line)
	respectively for the rising portion from 2002 to 2017
	and the part after the peak from 2017 to 2021.
{\it Bottom panel:} Detrended light curve, obtained from dividing 
the observed light curve by the best-fits to the underlying trends 
	shown in the top panel. The double-sinusoidal best-fit model 
	given by 
	the LSP code is shown as the yellow dashed curve.
\label{fig:lc}}
\end{figure}

We evaluated the significances of the two periodicities by simulating
the light curve. We used a smoothly bending power law
plus a constant, $P(f)=Af^{-\alpha}[1+(f/f_{bend})^{\beta - \alpha}]^{-1} +C$,
to model the underlying power spectrum density (PSD) of the flux density
variations \citep{gv12}. The model function consists of the parameters
$A$, $\alpha$, $\beta$, $f_{bend}$, and $C$, which are the normalization,
low frequency slope, high frequency slope, bend frequency, and Gaussian noise,
respectively.
A maximum likelihood method \citep{bv12} was used to fit the PSD, 
and we obtained
the best-fit values $A=0.27\pm0.17$, $\alpha=0.42\pm0.32$, $\beta= 0.59\pm0.29$,
$f_{bend}=0.012\pm0.109$, and $C=0.78\pm0.47$, respectively,
so for this source a single power-law would be acceptable.
Using the model, we generated $3\times 10^7$ light curves \citep{emm+13}
and  obtained the PSD of each. Significance curves were built by counting
the PSD data points at each frequency considered. The 
5$\sigma$ significance
curve obtained in this fashion is shown in Figure~\ref{fig:pow}. The two power peaks of 
the periodicities reach above the significance curve.
After considering the trial number in our temporal analysis, which is
350, $P_l$ and $P_h$ have only modestly reduced significances,
which respectively are greater than and close to 5$\sigma$.

The LSP code provides the best-fit model for the QPO, which
consists of two sinusoidal functions in a form of $ A\sin[2\pi f(t-t_0)]+A_0$.
The parameters for $P_l$ ($P_h$) are $A=1.47\pm0.13$ ($1.09\pm0.13$), 
$f=2.900\pm0.016\times 10^{-3}$\,d$^{-1}$ 
($2.585\pm0.020\times10^{-3}$\,d$^{-1}$), $t_0=137.51\pm4.68$ ($48.89\pm7.45$),
$A_0=5.37\pm0.09$ ($5.21\pm0.09$).
In order to clearly show the QPO variations in the 1.3-mm light curve, 
we detrended its long-term flux density changes.
The rising part before 2015 August 01 (of duration MJD~52577--57235)
was fitted with a degree 3 polynomial,  
$2.4-4.4\times 10^{-5}t + 5.2\times 10^{-7}t^2 -1.4\times 10^{-11}t^3$\,Jy
(with $t$ in days after MJD~52577).
The possible QPO 
variations, obtained by dividing the observed light curve with 
this function, are shown in Figure~\ref{fig:lc}. The normalized sinusoidal 
functions are overplotted. The model can approximately describe the
variations. In order to understand if the variations in the 2016--2021 part of
the light curve could be related, we extended this model fitting to the 
end of the light curve. The part corresponding to the 2016--2017 variation peak 
(MJD\,57235--57870) was still quite well fit by the polynomial, and because
the remaining part of 2017--2021 (MJD\,57870--59258) is relatively flat,
we fit it with a constant and normalized by the constant (9.18\,Jy).
The data points in general do not show a clear agreement with 
the double-sinusoidal modulation. The comparison may explain the analysis 
result that
the inclusion of the 2016--2021 light curve part do not help increase 
the significance.

As the SMA observations are prohibited from pointing to targets close to 
the Sun and thus the light curve we analyzed has 
a certain pattern that might induce artifactual signals \citep{van18}, 
we conducted detailed analyses to confirm 
the real presence of the QPO signal. The analyses are described 
in Appendix section~\ref{sec:a3}.
We did not find any evidence that the signal 
was an artifact.

\section{Discussion}
\label{sec:dis}

The QPO signal we have found is the first time one has been seen at mm 
wavelengths (to our knowledge) among the QPOs reported in emission at different 
wavelengths from Active Galactic Nuclei (AGN). While the radio and \gr\
light curves show similar long-term variation patterns,
we did not detect significant signals in them, in particular, the  
15\,GHz one (Figure~\ref{fig:lcs}). In this case, 
the mm observations may be able to more sensitively catch 
the variations 
of the jet's emission \citep{hod+18}, thus revealing the QPO signal.

Also uniquely, the QPO signal consists of two close periodicities 
and thus presents a very intriguing case.  The thus-far reported AGN QPOs 
are widely discussed to reflect instances such as periodic variations of 
the accretion disk surrounding a central SMBH or the orbital period of a 
binary SMBH (see, e.g., \citealt{gie+08,kin+13,ack+15,zw21}), and 
often compared to relatively well-studied cases in 
stellar-mass black hole systems as well.  In the latter, QPOs are seen
to appear as a series of harmonically related peaks \citep{im20}, $\nu_0$, 
2$\nu_0$, ..., (n+1)$\nu_0$, where $\nu_0$ is the fundamental peak frequency.
The double periods in our case obviously do not fit in this type of 
QPO phenomena, as $P_l$ and $P_h$ correspond to frequencies 
of 3.36$\times10^{-8}$\,Hz and 3.00$\times10^{-8}$\,Hz respectively. 
Theoretical simulations have shown that a thick disk in an AGN surrounding 
a SMBH may undergo acoustic p-mode oscillations when excited by an external 
perturbation \citep{rl05}. Such an oscillation signal could consist
of multiple frequencies, which may have been seen in a couple of periodic variability cases found in 
AGN (e.g., \citealt{an+13}).  However, the dominant frequencies should follow the 
ratio of 2:3, which is not the case here.

The disk/jet precession scenario due to the Lense-Thirring effect
(or frame-dragging effect) has been applied to the explanations of
QPO phenomena in X-ray binaries 
(e.g., \citealt{sm98,im20}), tidal disruption events (e.g., \citealt{pas+19}),
and blazars (e.g., \citealt{bd20}).
In this scenario, a thick disk around a rapidly spinning SMBH could be tilted 
and such a tilted disk would precess at a timescale of 
$\sim 0.02 (M/10^8 M_{\sun})(r/r_g)^3$\,d \citep{mer+10,bd20}, where
$M$ and $r_g =GM/c^2$ are the mass and the gravitational radius, 
respectively, of the SMBH.  For the NGC~1275 case, 
$M=3.4\times10^8\ M_{\sun}$ \citep{wej05}, and if the inner part of its presumed tilted disk 
is at $\sim20\ r_g$, the timescale would be $\sim 1$\,yr, possibly leading 
to similarly precessing jets \citep{lisk+18}.  So the QPO signal we have found to 
arise from
the jets of NGC~1275 could be another case of this scenario. However, there
are two close periodicities in our case. We note that in recent general 
relativistic 
magnetohydrodynamic simulations \citep{lisk+18}, the jets were found to be 
wobbling around the disk at an approximate 
timescale of $10^4r_g/c$, which is $\sim 0.5$\,yr 
for the case of NGC~1275. 
If we consider that  $P_l$ arises from the jets and is due to
wobbling, the disk precession timescale, $P_{\rm prec}$, could be much longer 
(and stable), resulting in the weaker periodicity $P_h$ through a beat frequency where
$P_{\rm prec}=1/(1/344.6\,{\rm d}- 1/386.0\,{\rm d}) \simeq 9$\,yr.
This $P_{\rm prec}$ value would put the inner part of the disk in NGC~1275 
at $\sim40 r_g$. This scenario, with a jet wobbling at $P_l$ and precessing
at $P_{\rm prec}$, is thus a possible way to explain the observed two periodicities.  Whether such a  wobbling 
jet could induce a significant QPO signal is not clear
and should be further investigated.

In addition, we note the well-studied case of 
the stellar-mass black hole binary system SS~433, which has an orbital 
period of $\sim 13$\,d and whose jets are precessing with a period 
of $\sim 162$\,d \citep{mar84}. Quite a few 
X-ray binaries, including SS~433, show a period much longer than the known 
orbital period, and this is considered to possibly reflect a precessing disk
that surrounds the central compact star in such an X-ray binary \citep{wp99}. 
So we could consider $P_l$ 
as an orbital period of a binary SMBH system and 
take the jets to be 
precessing at $P_{\rm prec}$ estimated above. 
The ratio of $P_{\rm prec}$ to $P_l$
would be $\simeq 9.3$, similar to that 
(162\,d/13\,d$\simeq 12.5$) in SS~433. Such ratios of other X-ray binaries
are generally in a range of 10--100 \citep{wp99}. 
This similarity leads to the possible scenario where the galaxy contains 
jets precessing with
a $\sim 9$\,yr period and that the center could consist of two SMBHs
orbiting each other at a period of $\sim 345$\,d.

There is evidence  that \ngc\ experienced a 
merger with another galaxy probably $\sim$300\,Myr ago \citep{car+98,con+01}. 
A binary SMBH system would
be a natural result of the merger. However current studies show that it often
takes giga-year for the two SMBHs from each galaxy to form a compact binary
after a merger event (e.g., \citealt{yu02}). Considering 345\,d as the 
orbital period,
the binary would have a short separation distance of $\sim$0.003\,pc 
(here assuming equal masses for the two SMBHs, with $M_1=M_2$ and 
$M_1+M_2= 3.4\times 10^8\,M_{\odot}$). Given the compactness \citep{yu02},
either this putative binary SMBH might not 
be the result of the previous merger but may have formed from an earlier merger,
or the formation of a compact SMBH binary could simply be a factor of 
a few faster than theoretically expected.

As shown by \citet{hod+18}, the C1 component was brighter and more
variable at 2--3\,mm wavelengths than the C3 component before the apparent 2015 August
C3 collision with a clump of gas. Thus the QPO signal we have found could very well arise 
from C1. Indeed, by analyzing
the position angle changes of C1 at 43\,GHz, \citet{dom+21} found a 
$\sim 29$\,yr precession or a $\sim 21$\,yr precession plus a $\sim 3.4$\,yr 
nutation. Although these possible periods do not match those of the QPO, it should be noted
that their observations were at radio and over a limited time span of 
9\,yr. As mm observations may be more attuned to catching the variations of 
the jets in \ngc, long-term mm high-resolution imaging of the jets 
would potentially be able to determine the precession period (or periods), 
verify our QPO results, and confidently identify the nature of the jet variation.


\begin{acknowledgments}

We thank the anonymous referee for helpful suggestions,
Y.\ Luo for discussions about binary supermassive black hole 
evolution, and F.\ Xie for discussions about QPOs in X-ray binaries.
The Submillimeter Array is a joint project between the Smithsonian 
Astrophysical Observatory and the Academia Sinica Institute of Astronomy 
and Astrophysics and is funded by the Smithsonian Institution and the 
Academia Sinica.  This research has made use of data from the OVRO 40-m 
monitoring program which was supported in part by NASA grants NNX08AW31G, 
NNX11A043G and NNX14AQ89G, and NSF grants AST-0808050 and AST-1109911, 
and private funding from Caltech and the MPIfR.

This research is supported by the National Key R \& D Program of China 
under grant No. 2018YFA0404204, the joint foundation of Department 
of Science and Technology of Yunnan Province and Yunnan University 
$[$2018FY001 (-003)$]$,  and the National Natural
	Science Foundation of China (12163006, 11633007). 
	Z.W. acknowledges the support by the Original Innovation Program 
	of the Chinese Academy of Sciences (E085021002).

\end{acknowledgments}

%






\bibliographystyle{aasjournal}

\begin{thebibliography}{}
\expandafter\ifx\csname natexlab\endcsname\relax\def\natexlab#1{#1}\fi

\bibitem[{{Abdo} {et~al.}(2009){Abdo}, {Ackermann}, {Ajello}, {Asano},
  {Baldini}, {Ballet}, {Barbiellini}, {Bastieri}, {Baughman}, {Bechtol},
  {Bellazzini}, {Blandford}, {Bloom}, {Bonamente}, {Borgland}, {Bregeon},
  {Brez}, {Brigida}, {Bruel}, {Burnett}, {Caliandro}, {Cameron}, {Caraveo},
  {Casandjian}, {Cavazzuti}, {Cecchi}, {Celotti}, {Chekhtman}, {Cheung},
  {Chiang}, {Ciprini}, {Claus}, {Cohen-Tanugi}, {Colafrancesco}, {Cominsky},
  {Conrad}, {Costamante}, {Dermer}, {de Angelis}, {de Palma}, {Digel},
  {Donato}, {do Couto e Silva}, {Drell}, {Dubois}, {Dumora}, {Farnier},
  {Favuzzi}, {Finke}, {Focke}, {Frailis}, {Fukazawa}, {Funk}, {Fusco},
  {Gargano}, {Georganopoulos}, {Germani}, {Giebels}, {Giglietto}, {Giordano},
  {Glanzman}, {Grenier}, {Grondin}, {Grove}, {Guillemot}, {Guiriec},
  {Hanabata}, {Harding}, {Hartman}, {Hayashida}, {Hays}, {Hughes},
  {J{\'o}hannesson}, {Johnson}, {Johnson}, {Johnson}, {Kadler}, {Kamae},
  {Kanai}, {Katagiri}, {Kataoka}, {Kawai}, {Kerr}, {Kn{\"o}dlseder}, {Kuehn},
  {Kuss}, {Latronico}, {Lemoine-Goumard}, {Longo}, {Loparco}, {Lott},
  {Lovellette}, {Lubrano}, {Madejski}, {Makeev}, {Mazziotta}, {McEnery},
  {Meurer}, {Michelson}, {Mitthumsiri}, {Mizuno}, {Moiseev}, {Monte},
  {Monzani}, {Morselli}, {Moskalenko}, {Murgia}, {Nakamori}, {Nolan}, {Norris},
  {Nuss}, {Ohsugi}, {Omodei}, {Orlando}, {Ormes}, {Paneque}, {Panetta},
  {Parent}, {Pepe}, {Pesce-Rollins}, {Piron}, {Porter}, {Rain{\`o}}, {Razzano},
  {Reimer}, {Reimer}, {Reposeur}, {Ritz}, {Rodriguez}, {Romani}, {Ryde},
  {Sadrozinski}, {Sambruna}, {Sanchez}, {Sander}, {Sato}, {Parkinson},
  {Sgr{\`o}}, {Smith}, {Smith}, {Spandre}, {Spinelli}, {Starck}, {Strickman},
  {Strong}, {Suson}, {Tajima}, {Takahashi}, {Takahashi}, {Tanaka}, {Taylor},
  {Thayer}, {Thompson}, {Torres}, {Tosti}, {Uchiyama}, {Usher}, {Vilchez},
  {Vitale}, {Waite}, {Wood}, {Ylinen}, {Ziegler}, {Aller}, {Aller},
  {Kellermann}, {Kovalev}, {Kovalev}, {Lister}, \& {Pushkarev}}]{abd+09}
{Abdo}, A.~A., {Ackermann}, M., {Ajello}, M., {et~al.} 2009, \apj, 699, 31

\bibitem[{{Abdollahi} {et~al.}(2020){Abdollahi}, {Acero}, {Ackermann}, {Ajello
  }, {Atwood}, {Axelsson}, {Baldini}, {Ballet}, {Barbiellini}, {Bastieri},
  {Becerra Gonzalez}, {Bellazz ini}, {Berretta}, {Bissaldi}, {Blandford},
  {Bloom }, {Bonino}, {Bottacini}, {Brandt}, {Bregeon}, {Bruel}, {Buehler},
  {Burnett}, {Buson}, { Cameron}, {Caputo}, {Caraveo}, {Casandjian},
  {Cavazzuti}, {Charles}, {Chaty}, {Chen}, {Cheung}, {Chiaro}, {Ciprini},
  {Cohen-Tanugi}, {Cominsky}, {Coronado-Bl{\'a}zquez}, {Costantin}, {Cuoco},
  {Cutini}, {D'Ammando}, {DeKlotz}, {de la To rre Luque}, {de Palma}, {Desai},
  {Digel}, {Di Lal la}, {Di Mauro}, {Di Venere}, {Dom{\'\i}nguez}, {Dum ora},
  {Fana Dirirsa}, {Fegan}, {Ferrara}, {Fra nckowiak}, {Fukazawa}, {Funk},
  {Fusco}, {Gargano}, {Gasparrini}, {Giglietto}, {Giommi}, {Giordano},
  {Glanzman}, {Green}, {Grenier}, { Griffin}, {Grondin}, {Grove}, {Guiriec},
  {Har ding}, {Hayashi}, {Hays}, {Hewitt}, {Horan}, {J{\'o}hannesson},
  {Johnson}, {Kamae}, {Kerr}, {Kocevski}, {Kovac'evic'}, {Kuss}, {Landriu}, {L
  arsson}, {Latronico}, {Lemoine-Goumard}, {Li}, {Liodakis}, {Longo},
  {Loparco}, {Lott}, {Lovellette}, {Lubrano}, {Madejski}, {Maldera},
  {Malyshev}, {Manfreda}, {Marchesini}, {Marcotulli}, {Mart{\'\i}- Devesa},
  {Martin}, {Massaro}, {Mazziotta}, {McEne ry}, {Mereu}, {Meyer}, {Michelson},
  {Mirabal}, {Mizuno}, {Monzani}, {Morselli}, {Moskalenko}, {Negro}, {Nuss},
  {Ojha}, {Omodei}, {Orienti }, {Orlando}, {Ormes}, {Palatiello}, {Paliya},
  {Paneque}, {Pei}, {Pe{\~n}a-Herazo}, {Perkins}, {Persic}, {Pesce-Rollins},
  {Petrosian}, {Petrov}, {Piron}, {Poon}, {Porter}, {Principe}, {Rain {\`o}},
  {Rando}, {Razzano}, {Razzaque}, {Reimer}, {Reimer}, {Remy}, {Reposeur},
  {Romani}, {Saz Parkinson}, {Schinzel}, {Serini}, {Sgr{\`o}}, {Siskind},
  {Smith}, {Spandre}, {Spinelli}, {S trong}, {Suson}, {Tajima}, {Takahashi},
  {Ta k}, {Thayer}, {Thompson}, {Tibaldo}, {Torres}, {Torresi}, {Valverde},
  {Van Klaveren}, {van Zyl}, {Wood}, {Yassine}, \& {Zaharijas}}]{abd+20}
{Abdollahi}, S., {Acero}, F., {Ackermann}, M., {et~al.} 2020, \apjs, 247, 33

\bibitem[{{Ackermann} {et~al.}(2015){Ackermann}, {Ajello}, {Albert}, {Atwood},
  {Baldini}, {Ballet}, {Barbiellini}, {Bastieri}, {Becerra Gonzalez},
  {Bellazzini}, {Bissaldi}, {Blandford}, {Bloom}, {Bonino}, {Bottacini},
  {Bregeon}, {Bruel}, {Buehler}, {Buson}, {Caliandro}, {Cameron}, {Caputo},
  {Caragiulo}, {Caraveo}, {Cavazzuti}, {Cecchi}, {Chekhtman}, {Chiang},
  {Chiaro}, {Ciprini}, {Cohen-Tanugi}, {Conrad}, {Cutini}, {D'Ammando}, {de
  Angelis}, {de Palma}, {Desiante}, {Di Venere}, {Dom{\'\i}nguez}, {Drell},
  {Favuzzi}, {Fegan}, {Ferrara}, {Focke}, {Fuhrmann}, {Fukazawa}, {Fusco},
  {Gargano}, {Gasparrini}, {Giglietto}, {Giommi}, {Giordano}, {Giroletti},
  {Godfrey}, {Green}, {Grenier}, {Grove}, {Guiriec}, {Harding}, {Hays},
  {Hewitt}, {Hill}, {Horan}, {Jogler}, {J{\'o}hannesson}, {Johnson}, {Kamae},
  {Kuss}, {Larsson}, {Latronico}, {Li}, {Li}, {Longo}, {Loparco}, {Lott},
  {Lovellette}, {Lubrano}, {Magill}, {Maldera}, {Manfreda}, {Max-Moerbeck},
  {Mayer}, {Mazziotta}, {McEnery}, {Michelson}, {Mizuno}, {Monzani},
  {Morselli}, {Moskalenko}, {Murgia}, {Nuss}, {Ohno}, {Ohsugi}, {Ojha},
  {Omodei}, {Orlando}, {Ormes}, {Paneque}, {Pearson}, {Perkins}, {Perri},
  {Pesce-Rollins}, {Petrosian}, {Piron}, {Pivato}, {Porter}, {Rain{\`o}},
  {Rando}, {Razzano}, {Readhead}, {Reimer}, {Reimer}, {Schulz}, {Sgr{\`o}},
  {Siskind}, {Spada}, {Spandre}, {Spinelli}, {Suson}, {Takahashi}, {Thayer},
  {Thompson}, {Tibaldo}, {Torres}, {Tosti}, {Troja}, {Uchiyama}, {Vianello},
  {Wood}, {Wood}, {Zimmer}, {Berdyugin}, {Corbet}, {Hovatta}, {Lindfors},
  {Nilsson}, {Reinthal}, {Sillanp{\"a}{\"a}}, {Stamerra}, {Takalo}, \&
  {Valtonen}}]{ack+15}
{Ackermann}, M., {Ajello}, M., {Albert}, A., {et~al.} 2015, \apjl, 813, L41

\bibitem[{{Aleksi{\'c}} {et~al.}(2012){Aleksi{\'c}}, {Alvarez}, {Antonelli},
  {Antoranz}, {Asensio}, {Backes}, {Barres de Almeida}, {Barrio}, {Bastieri},
  {Becerra Gonz{\'a}lez}, {Bednarek}, {Berger}, {Bernardini}, {Biland},
  {Blanch}, {Bock}, {Boller}, {Bonnoli}, {Borla Tridon}, {Bretz},
  {Ca{\~n}ellas}, {Carmona}, {Carosi}, {Colin}, {Colombo}, {Contreras},
  {Cortina}, {Cossio}, {Covino}, {da Vela}, {Dazzi}, {de Angelis}, {de Caneva},
  {de Cea Del Pozo}, {de Lotto}, {Delgado Mendez}, {Diago Ortega}, {Doert},
  {Dom{\'\i}nguez}, {Dominis Prester}, {Dorner}, {Doro}, {Eisenacher},
  {Elsaesser}, {Ferenc}, {Fonseca}, {Font}, {Fruck}, {Garc{\'\i}a L{\'o}pez},
  {Garczarczyk}, {Garrido}, {Giavitto}, {Godinovi{\'c}}, {Gozzini}, {Hadasch},
  {H{\"a}fner}, {Herrero}, {Hildebrand}, {H{\"o}hne-M{\"o}nch}, {Hose},
  {Hrupec}, {Huber}, {Jogler}, {Kadenius}, {Kellermann}, {Klepser},
  {Kr{\"a}henb{\"u}hl}, {Krause}, {La Barbera}, {Lelas}, {Leonardo},
  {Lewandowska}, {Lindfors}, {Lombardi}, {L{\'o}pez}, {L{\'o}pez-Coto},
  {L{\'o}pez-Oramas}, {Lorenz}, {Makariev}, {Maneva}, {Mankuzhiyil},
  {Mannheim}, {Maraschi}, {Mariotti}, {Mart{\'\i}nez}, {Mazin}, {Meucci},
  {Miranda}, {Mirzoyan}, {Mold{\'o}n}, {Moralejo}, {Munar-Adrover},
  {Niedzwiecki}, {Nieto}, {Nilsson}, {Nowak}, {Orito}, {Paiano}, {Paneque},
  {Paoletti}, {Pardo}, {Paredes}, {Partini}, {Perez-Torres}, {Persic},
  {Peruzzo}, {Pilia}, {Pochon}, {Prada}, {Prada Moroni}, {Prandini}, {Puerto
  Gimenez}, {Puljak}, {Reichardt}, {Reinthal}, {Rhode}, {Rib{\'o}}, {Rico},
  {R{\"u}gamer}, {Saggion}, {Saito}, {Saito}, {Salvati}, {Satalecka},
  {Scalzotto}, {Scapin}, {Schultz}, {Schweizer}, {Shayduk}, {Shore},
  {Sillanp{\"a}{\"a}}, {Sitarek}, {Snidaric}, {Sobczynska}, {Spanier}, {Spiro},
  {Stamatescu}, {Stamerra}, {Steinke}, {Storz}, {Strah}, {Sun}, {Suri{\'c}},
  {Takalo}, {Takami}, {Tavecchio}, {Temnikov}, {Terzi{\'c}}, {Tescaro},
  {Teshima}, {Tibolla}, {Torres}, {Treves}, {Uellenbeck}, {Vogler}, {Wagner},
  {Weitzel}, {Zabalza}, {Zandanel}, {Zanin}, {Pfrommer}, \& {Pinzke}}]{ale+12}
{Aleksi{\'c}}, J., {Alvarez}, E.~A., {Antonelli}, L.~A., {et~al.} 2012, \aap,
  539, L2

\bibitem[{{An} {et~al.}(2013){An}, {Baan}, {Wang}, {Wang}, \& {Hong}}]{an+13}
{An}, T., {Baan}, W.~A., {Wang}, J.-Y., {Wang}, Y., \& {Hong}, X.-Y. 2013,
  \mnras, 434, 3487

\bibitem[{{Atwood} {et~al.}(2009){Atwood}, {Abdo}, {Ackermann}, {Althouse},
  {Anderson}, {Axelsson}, {Baldini}, {Ballet}, {Band}, {Barbiellini},
  {Bartelt}, {Bastieri}, {Baughman}, {Bechtol}, {B{\'e}d{\'e}r{\`e}de},
  {Bellardi}, {Bellazzini}, {Berenji}, {Bignami}, {Bisello}, {Bissaldi},
  {Blandford}, {Bloom}, {Bogart}, {Bonamente}, {Bonnell}, {Borgland},
  {Bouvier}, {Bregeon}, {Brez}, {Brigida}, {Bruel}, {Burnett}, {Busetto},
  {Caliandro}, {Cameron}, {Caraveo}, {Carius}, {Carlson}, {Casandjian},
  {Cavazzuti}, {Ceccanti}, {Cecchi}, {Charles}, {Chekhtman}, {Cheung},
  {Chiang}, {Chipaux}, {Cillis}, {Ciprini}, {Claus}, {Cohen-Tanugi},
  {Condamoor}, {Conrad}, {Corbet}, {Corucci}, {Costamante}, {Cutini}, {Davis},
  {Decotigny}, {DeKlotz}, {Dermer}, {de Angelis}, {Digel}, {do Couto e Silva},
  {Drell}, {Dubois}, {Dumora}, {Edmonds}, {Fabiani}, {Farnier}, {Favuzzi},
  {Flath}, {Fleury}, {Focke}, {Funk}, {Fusco}, {Gargano}, {Gasparrini},
  {Gehrels}, {Gentit}, {Germani}, {Giebels}, {Giglietto}, {Giommi}, {Giordano},
  {Glanzman}, {Godfrey}, {Grenier}, {Grondin}, {Grove}, {Guillemot}, {Guiriec},
  {Haller}, {Harding}, {Hart}, {Hays}, {Healey}, {Hirayama}, {Hjalmarsdotter},
  {Horn}, {Hughes}, {J{\'o}hannesson}, {Johansson}, {Johnson}, {Johnson},
  {Johnson}, {Johnson}, {Kamae}, {Katagiri}, {Kataoka}, {Kavelaars}, {Kawai},
  {Kelly}, {Kerr}, {Klamra}, {Kn{\"o}dlseder}, {Kocian}, {Komin}, {Kuehn},
  {Kuss}, {Landriu}, {Latronico}, {Lee}, {Lee}, {Lemoine-Goumard}, {Lionetto},
  {Longo}, {Loparco}, {Lott}, {Lovellette}, {Lubrano}, {Madejski}, {Makeev},
  {Marangelli}, {Massai}, {Mazziotta}, {McEnery}, {Menon}, {Meurer},
  {Michelson}, {Minuti}, {Mirizzi}, {Mitthumsiri}, {Mizuno}, {Moiseev},
  {Monte}, {Monzani}, {Moretti}, {Morselli}, {Moskalenko}, {Murgia},
  {Nakamori}, {Nishino}, {Nolan}, {Norris}, {Nuss}, {Ohno}, {Ohsugi}, {Omodei},
  {Orlando}, {Ormes}, {Paccagnella}, {Paneque}, {Panetta}, {Parent}, {Pearce},
  {Pepe}, {Perazzo}, {Pesce-Rollins}, {Picozza}, {Pieri}, {Pinchera}, {Piron},
  {Porter}, {Poupard}, {Rain{\`o}}, {Rando}, {Rapposelli}, {Razzano}, {Reimer},
  {Reimer}, {Reposeur}, {Reyes}, {Ritz}, {Rochester}, {Rodriguez}, {Romani},
  {Roth}, {Russell}, {Ryde}, {Sabatini}, {Sadrozinski}, {Sanchez}, {Sander},
  {Sapozhnikov}, {Parkinson}, {Scargle}, {Schalk}, {Scolieri}, {Sgr{\`o}},
  {Share}, {Shaw}, {Shimokawabe}, {Shrader}, {Sierpowska-Bartosik}, {Siskind},
  {Smith}, {Smith}, {Spandre}, {Spinelli}, {Starck}, {Stephens}, {Strickman},
  {Strong}, {Suson}, {Tajima}, {Takahashi}, {Takahashi}, {Tanaka}, {Tenze},
  {Tether}, {Thayer}, {Thayer}, {Thompson}, {Tibaldo}, {Tibolla}, {Torres},
  {Tosti}, {Tramacere}, {Turri}, {Usher}, {Vilchez}, {Vitale}, {Wang},
  {Watters}, {Winer}, {Wood}, {Ylinen}, \& {Ziegler}}]{atw+09}
{Atwood}, W.~B., {Abdo}, A.~A., {Ackermann}, M., {et~al.} 2009, \apj, 697, 1071

\bibitem[{{Barret} \& {Vaughan}(2012)}]{bv12}
{Barret}, D., \& {Vaughan}, S. 2012, \apj, 746, 131

\bibitem[{{Bhatta} \& {Dhital}(2020)}]{bd20}
{Bhatta}, G., \& {Dhital}, N. 2020, \apj, 891, 120

\bibitem[{{Britzen} {et~al.}(2019){Britzen}, {Fendt}, {Zaja{\v{c}}ek}, {Jaron},
  {Pashchenko}, {Aller}, \& {Aller}}]{bri+19}
{Britzen}, S., {Fendt}, C., {Zaja{\v{c}}ek}, M., {et~al.} 2019, Galaxies, 7, 72

\bibitem[{{Carlson} {et~al.}(1998){Carlson}, {Holtzman}, {Watson}, {Grillmair},
  {Mould}, {Ballester}, {Burrows}, {Clarke}, {Crisp}, {Evans}, {Gallagher},
  {Griffiths}, {Hester}, {Hoessel}, {Scowen}, {Stapelfeldt}, {Trauger}, \&
  {Westphal}}]{car+98}
{Carlson}, M.~N., {Holtzman}, J.~A., {Watson}, A.~M., {et~al.} 1998, \aj, 115,
  1778

\bibitem[{{Conselice} {et~al.}(2001){Conselice}, {Gallagher}, \&
  {Wyse}}]{con+01}
{Conselice}, C.~J., {Gallagher}, John~S., I., \& {Wyse}, R. F.~G. 2001, \aj,
  122, 2281

\bibitem[{{Dominik} {et~al.}(2021){Dominik}, {Linhoff}, {Els{\"a}sser}, \&
  {Rhode}}]{dom+21}
{Dominik}, R.~M., {Linhoff}, L., {Els{\"a}sser}, D., \& {Rhode}, W. 2021,
  \mnras, 503, 5448

\bibitem[{{Dunn} {et~al.}(2006){Dunn}, {Fabian}, \& {Sanders}}]{dfs06}
{Dunn}, R.~J.~H., {Fabian}, A.~C., \& {Sanders}, J.~S. 2006, \mnras, 366, 758

\bibitem[{{Emmanoulopoulos} {et~al.}(2013){Emmanoulopoulos}, {McHardy}, \&
  {Papadakis}}]{emm+13}
{Emmanoulopoulos}, D., {McHardy}, I.~M., \& {Papadakis}, I.~E. 2013, \mnras,
  433, 907

\bibitem[{{Foster}(1996)}]{fos96}
{Foster}, G. 1996, \aj, 112, 1709

\bibitem[{{Gierli{\'n}ski} {et~al.}(2008){Gierli{\'n}ski}, {Middleton}, {Ward},
  \& {Done}}]{gie+08}
{Gierli{\'n}ski}, M., {Middleton}, M., {Ward}, M., \& {Done}, C. 2008, \nat,
  455, 369

\bibitem[{{Gonz{\'a}lez-Mart{\'\i}n} \& {Vaughan}(2012)}]{gv12}
{Gonz{\'a}lez-Mart{\'\i}n}, O., \& {Vaughan}, S. 2012, \aap, 544, A80

\bibitem[{{Gulati} {et~al.}(2021){Gulati}, {Bhattacharya}, {Bhattacharyya},
  {Bhatt}, {Stalin}, \& {Agrawal}}]{gul+21}
{Gulati}, S., {Bhattacharya}, D., {Bhattacharyya}, S., {et~al.} 2021, \mnras,
  503, 446

\bibitem[{{Gurwell} {et~al.}(2007){Gurwell}, {Peck}, {Hostler}, {Darrah}, \&
  {Katz}}]{gur+07}
{Gurwell}, M.~A., {Peck}, A.~B., {Hostler}, S.~R., {Darrah}, M.~R., \& {Katz},
  C.~A. 2007, in Astronomical Society of the Pacific Conference Series, Vol.
  375, From Z-Machines to ALMA: (Sub)Millimeter Spectroscopy of Galaxies, ed.
  A.~J. {Baker}, J.~{Glenn}, A.~I. {Harris}, J.~G. {Mangum}, \& M.~S. {Yun},
  234

\bibitem[{{Hiura} {et~al.}(2018){Hiura}, {Nagai}, {Kino}, {Niinuma}, {Sorai},
  {Chida}, {Akiyama}, {D'Ammando}, {Giovannini}, {Giroletti}, {Hada}, {Honma},
  {Koyama}, {Orienti}, {Orosz}, \& {Sawada-Satoh}}]{hiu+18}
{Hiura}, K., {Nagai}, H., {Kino}, M., {et~al.} 2018, \pasj, 70, 83

\bibitem[{{Hodgson} {et~al.}(2018){Hodgson}, {Rani}, {Lee}, {Algaba}, {Kino},
  {Trippe}, {Park}, {Zhao}, {Byun}, {Kang}, {Kim}, {Kim}, {Kim}, {Miyazaki},
  {Wajima}, {Oh}, {Kim}, \& {Gurwell}}]{hod+18}
{Hodgson}, J.~A., {Rani}, B., {Lee}, S.-S., {et~al.} 2018, \mnras, 475, 368

\bibitem[{{Ingram} \& {Motta}(2019)}]{im20}
{Ingram}, A.~R., \& {Motta}, S.~E. 2019, \nar, 85, 101524

\bibitem[{{King} {et~al.}(2013){King}, {Hovatta}, {Max-Moerbeck}, {Meier},
  {Pearson}, {Readhead}, {Reeves}, {Richards}, \& {Shepherd}}]{kin+13}
{King}, O.~G., {Hovatta}, T., {Max-Moerbeck}, W., {et~al.} 2013, \mnras, 436,
  L114

\bibitem[{{Kino} {et~al.}(2018){Kino}, {Wajima}, {Kawakatu}, {Nagai},
  {Orienti}, {Giovannini}, {Hada}, {Niinuma}, \& {Giroletti}}]{kin+18}
{Kino}, M., {Wajima}, K., {Kawakatu}, N., {et~al.} 2018, \apj, 864, 118

\bibitem[{{Liska} {et~al.}(2018){Liska}, {Hesp}, {Tchekhovskoy}, {Ingram}, {van
  der Klis}, \& {Markoff}}]{lisk+18}
{Liska}, M., {Hesp}, C., {Tchekhovskoy}, A., {et~al.} 2018, \mnras, 474, L81

\bibitem[{{Lister} {et~al.}(2018){Lister}, {Aller}, {Aller}, {Hodge}, {Homan},
  {Kovalev}, {Pushkarev}, \& {Savolainen}}]{lis+18}
{Lister}, M.~L., {Aller}, M.~F., {Aller}, H.~D., {et~al.} 2018, \apjs, 234, 12

\bibitem[{{Lomb}(1976)}]{lom76}
{Lomb}, N.~R. 1976, \apss, 39, 447

\bibitem[{{Margon}(1984)}]{mar84}
{Margon}, B. 1984, \araa, 22, 507

\bibitem[{{Merritt} {et~al.}(2010){Merritt}, {Alexander}, {Mikkola}, \&
  {Will}}]{mer+10}
{Merritt}, D., {Alexander}, T., {Mikkola}, S., \& {Will}, C.~M. 2010, \prd, 81,
  062002

\bibitem[{{Nagai} {et~al.}(2010){Nagai}, {Suzuki}, {Asada}, {Kino}, {Kameno},
  {Doi}, {Inoue}, {Kataoka}, {Bach}, {Hirota}, {Matsumoto}, {Honma},
  {Kobayashi}, \& {Fujisawa}}]{nag+10}
{Nagai}, H., {Suzuki}, K., {Asada}, K., {et~al.} 2010, \pasj, 62, L11

\bibitem[{{Nemmen} {et~al.}(2020){Nemmen}, {de Menezes}, \&
  {Paschalidis}}]{ndp20}
{Nemmen}, R., {de Menezes}, R., \& {Paschalidis}, V. 2020, in Perseus in
  Sicily: From Black Hole to Cluster Outskirts, ed. K.~{Asada}, E.~{de Gouveia
  Dal Pino}, M.~{Giroletti}, H.~{Nagai}, \& R.~{Nemmen}, Vol. 342, 167--171

\bibitem[{{Pasham} {et~al.}(2019){Pasham}, {Remillard}, {Fragile}, {Franchini},
  {Stone}, {Lodato}, {Homan}, {Chakrabarty}, {Baganoff}, {Steiner}, {Coughlin},
  \& {Pasham}}]{pas+19}
{Pasham}, D.~R., {Remillard}, R.~A., {Fragile}, P.~C., {et~al.} 2019, Science,
  363, 531

\bibitem[{{Readhead} {et~al.}(1989){Readhead}, {Lawrence}, {Myers}, {Sargent},
  {Hardebeck}, \& {Moffet}}]{rea+89}
{Readhead}, A.~C.~S., {Lawrence}, C.~R., {Myers}, S.~T., {et~al.} 1989, \apj,
  346, 566

\bibitem[{{Richards} {et~al.}(2011){Richards}, {Max-Moerbeck}, {Pavlidou},
  {King}, {Pearson}, {Readhead}, {Reeves}, {Shepherd}, {Stevenson},
  {Weintraub}, {Fuhrmann}, {Angelakis}, {Zensus}, {Healey}, {Romani}, {Shaw},
  {Grainge}, {Birkinshaw}, {Lancaster}, {Worrall}, {Taylor}, {Cotter}, \&
  {Bustos}}]{ric+11}
{Richards}, J.~L., {Max-Moerbeck}, W., {Pavlidou}, V., {et~al.} 2011, \apjs,
  194, 29

\bibitem[{{Rubio-Herrera} \& {Lee}(2005)}]{rl05}
{Rubio-Herrera}, E., \& {Lee}, W.~H. 2005, \mnras, 357, L31

\bibitem[{{Scargle}(1982)}]{sca82}
{Scargle}, J.~D. 1982, \apj, 263, 835

\bibitem[{{Stella} \& {Vietri}(1998)}]{sm98}
{Stella}, L., \& {Vietri}, M. 1998, \apjl, 492, L59

\bibitem[{{Suzuki} {et~al.}(2012){Suzuki}, {Nagai}, {Kino}, {Kataoka}, {Asada},
  {Doi}, {Inoue}, {Orienti}, {Giovannini}, {Giroletti}, {L{\"a}hteenm{\"a}ki},
  {Tornikoski}, {Le{\'o}n-Tavares}, {Bach}, {Kameno}, \& {Kobayashi}}]{suz+12}
{Suzuki}, K., {Nagai}, H., {Kino}, M., {et~al.} 2012, \apj, 746, 140

\bibitem[{{VanderPlas}(2018)}]{van18}
{VanderPlas}, J.~T. 2018, \apjs, 236, 16

\bibitem[{{Vermeulen} {et~al.}(1994){Vermeulen}, {Readhead}, \&
  {Backer}}]{vrb94}
{Vermeulen}, R.~C., {Readhead}, A.~C.~S., \& {Backer}, D.~C. 1994, \apjl, 430,
  L41

\bibitem[{{V{\'e}ron-Cetty} \& {V{\'e}ron}(2010)}]{vv10}
{V{\'e}ron-Cetty}, M.~P., \& {V{\'e}ron}, P. 2010, \aap, 518, A10

\bibitem[{{Wijers} \& {Pringle}(1999)}]{wp99}
{Wijers}, R. A.~M.~J., \& {Pringle}, J.~E. 1999, \mnras, 308, 207

\bibitem[{{Wilman} {et~al.}(2005){Wilman}, {Edge}, \& {Johnstone}}]{wej05}
{Wilman}, R.~J., {Edge}, A.~C., \& {Johnstone}, R.~M. 2005, \mnras, 359, 755

\bibitem[{{Yu}(2002)}]{yu02}
{Yu}, Q. 2002, \mnras, 331, 935

\bibitem[{{Zechmeister} \& {K{\"u}rster}(2009)}]{zk09}
{Zechmeister}, M., \& {K{\"u}rster}, M. 2009, \aap, 496, 577

\bibitem[{Zhang \& Wang(2021)}]{zw21}
Zhang, P., \& Wang, Z. 2021, ApJ, 914, 1

\end{thebibliography}

\appendix
\label{sec:app}

\restartappendixnumbering

\section{Illustration of the jets in the nucleus of \ngc}
\label{sec:a1}

\begin{figure}
\begin{center}
\includegraphics[width=0.6\linewidth]{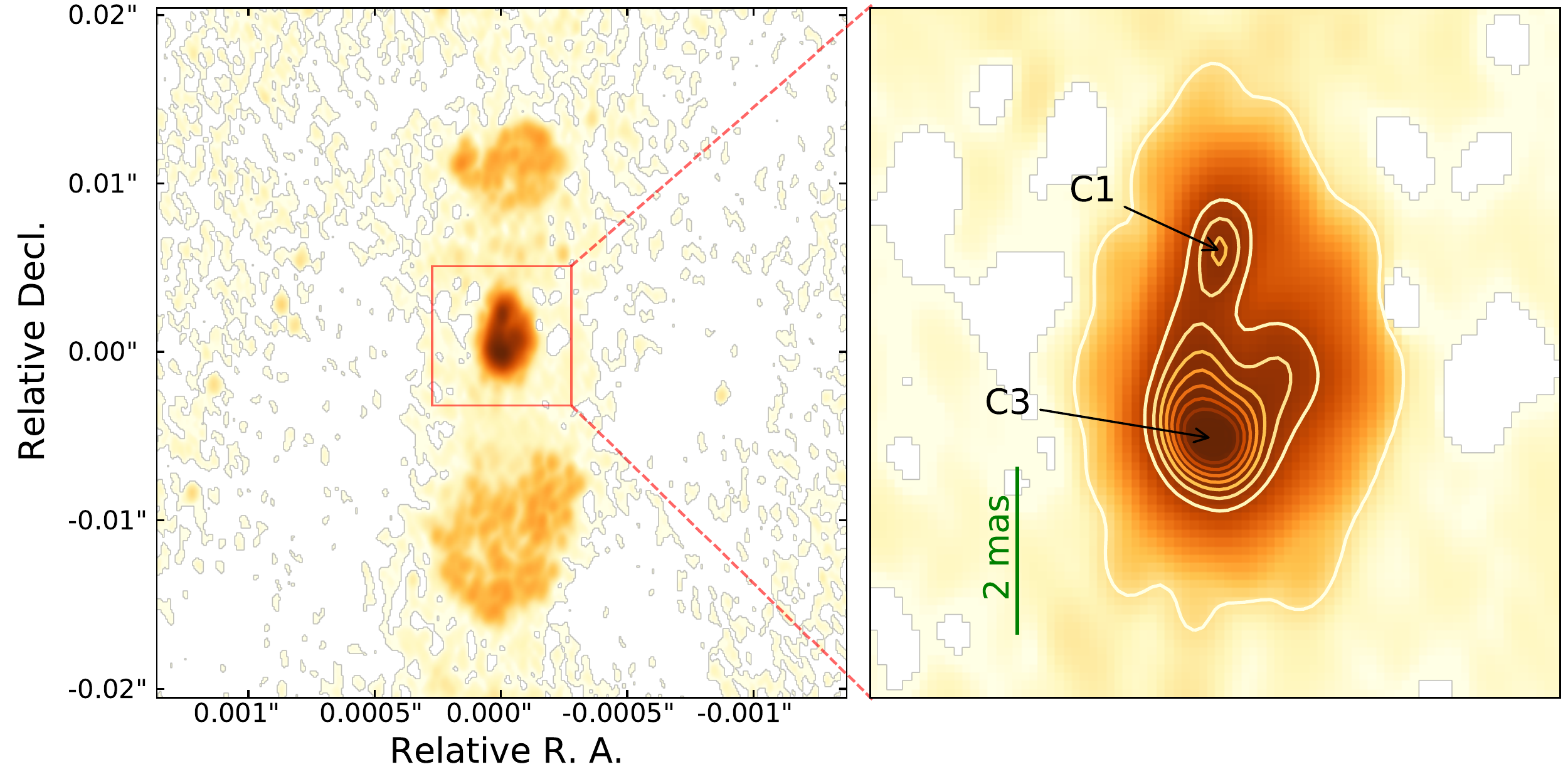}
\end{center}
	\caption{{\it Left:} 15\,GHz radio image 
	of the central 20 mas region of NGC~1275. 
	{\it Right:} enlarged view of 
	the central part, with
	the C1 and C3 components indicated. }
\label{fig:rimg}
\end{figure}

In Figure~\ref{fig:rimg}, the 15\,GHz radio image, taken in 2014 from
the Monitoring Of Jets in Active galactic nuclei with VLBA Experiments (MOJAVE;
\citealt{lis+18}) is shown. In the right panel of the figure, the C1 and C3
components are indicated.  Previously there was a relatively
	faint component C2, which was located 
	west (at the right side) of C3 \citep{suz+12}.

\section{Long-term light curves of NGC~1275 and periodicity analysis results}
\label{sec:a2}

\begin{figure*}
\centering
\epsscale{1.0}
	\includegraphics[width=0.43\linewidth]{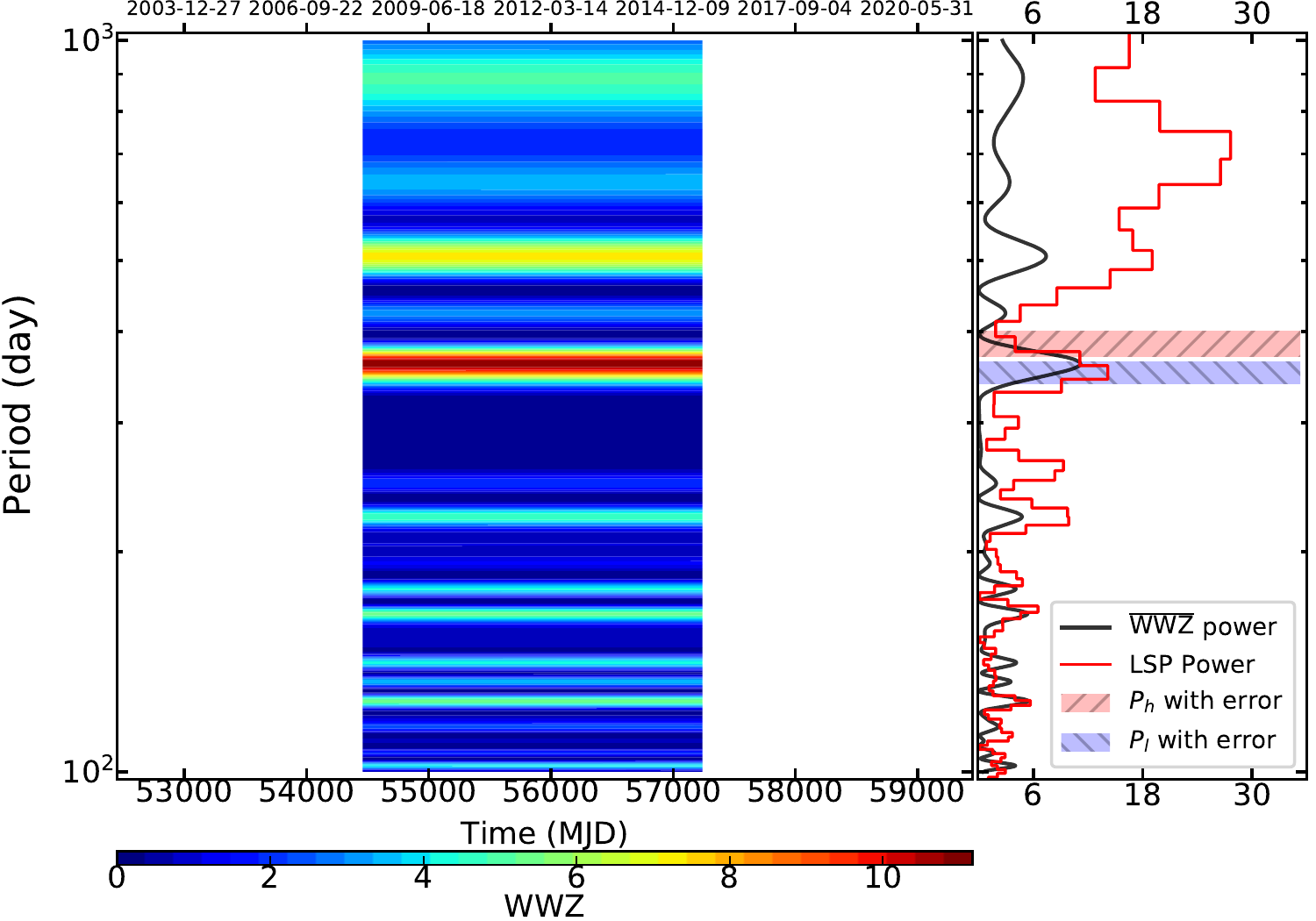}
	\includegraphics[width=0.43\linewidth]{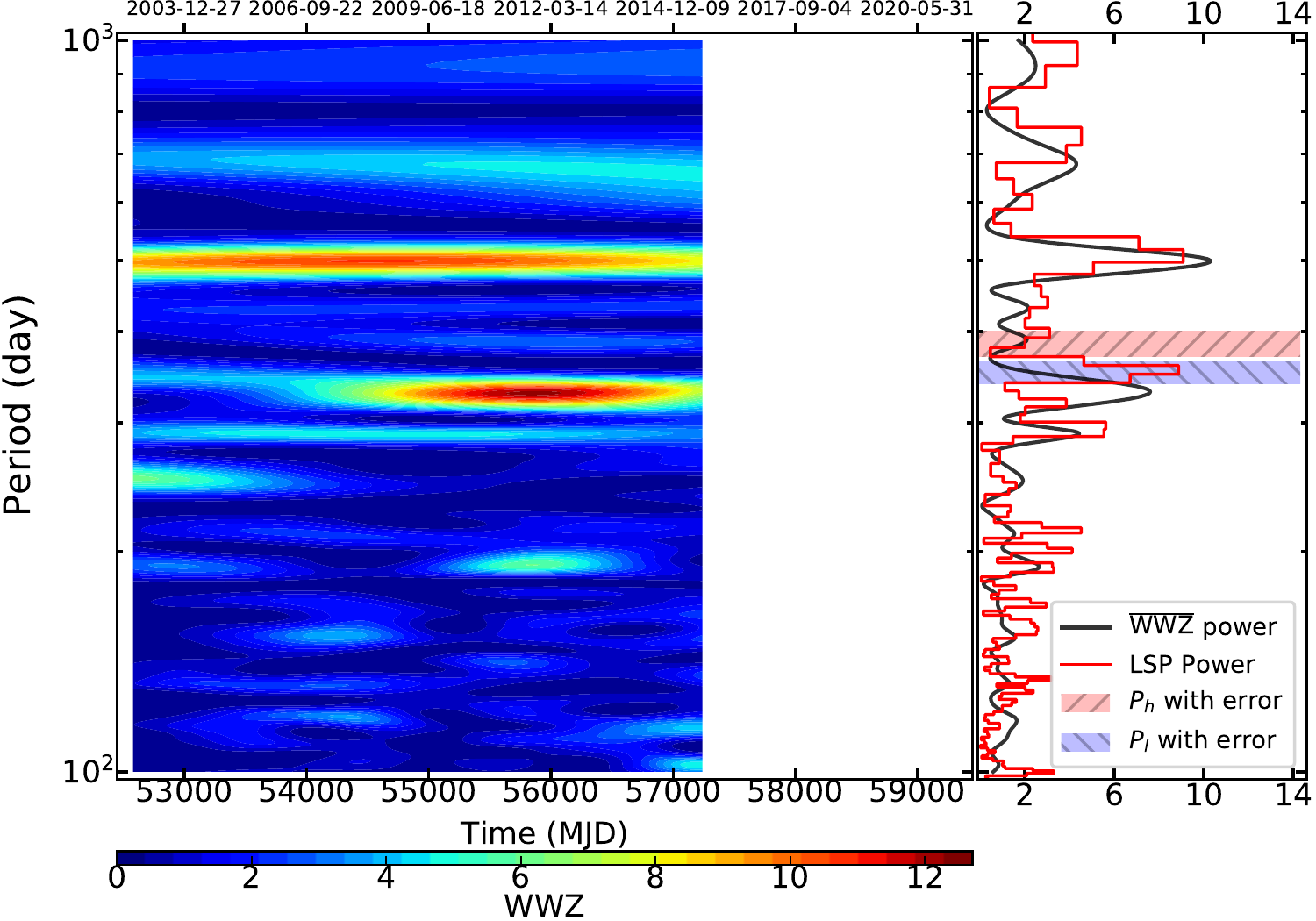}
	\vspace{15mm}
	\includegraphics[width=0.43\linewidth]{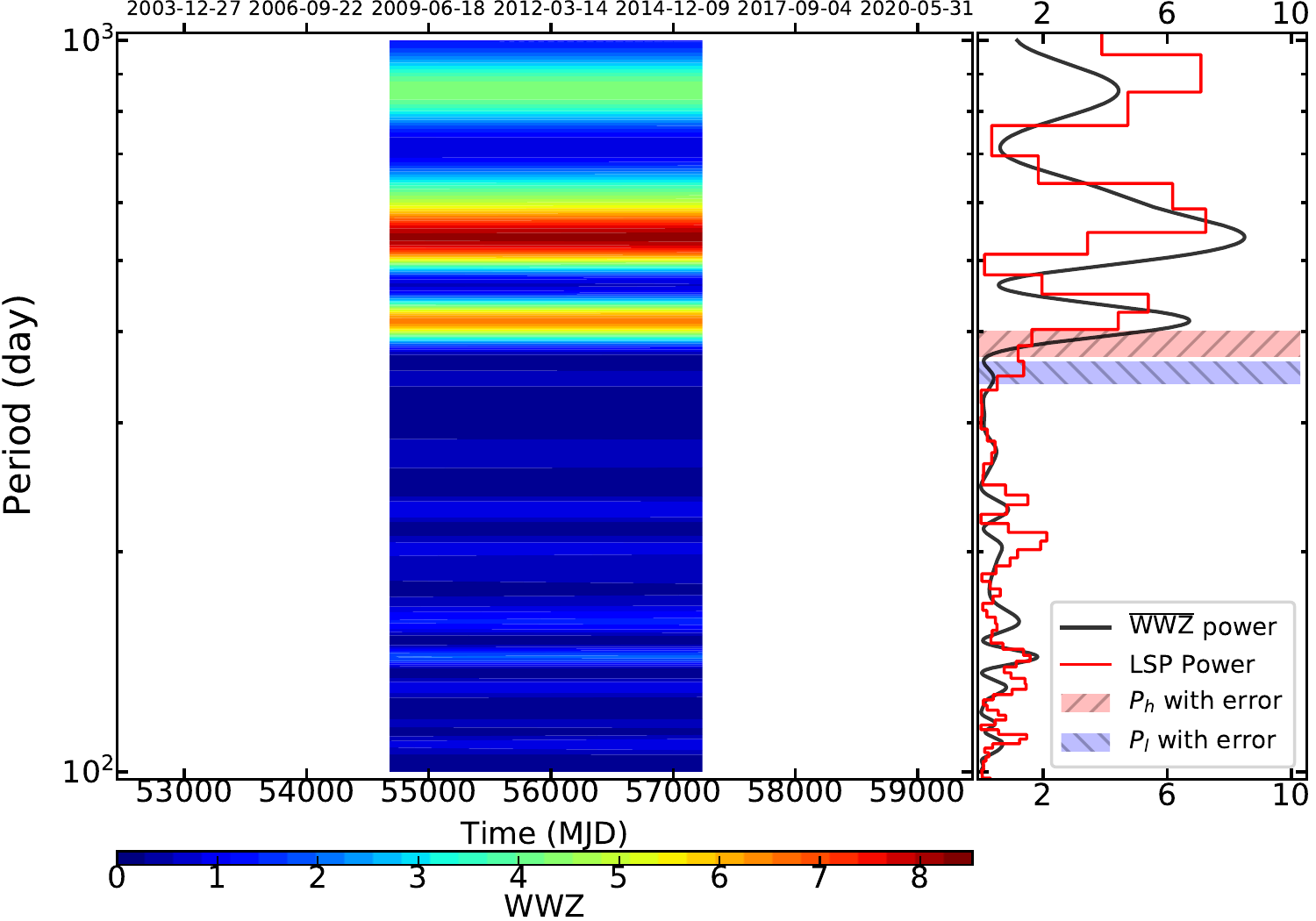}
	\caption{WWZ powers and LSP powers for the 
	15\,GHz radio ({\it top left}), 870-$\mu$m ({\it top right}), and
	{\it Fermi}-LAT 0.1--500\,GeV ({\it bottom}) light curves of NGC~1275. 
	For each of these bands there is one modest power peak 
	close to the periodicity ranges 
	determined from the 1.3-mm light curve (shown with pink and blue 
	shadings).  In addition, the 870~$\mu$m band shows
	a rather strong peak around 500\,d and the others have some signals 
	longer than 500\,d.
	However, considering the sparseness of the data points at 870\,$\mu$m and
	high values of each underlying power spectral density at large periods, none of
	the signals at 870-$\mu$m or $\gamma$-rays are significant.  
\label{fig:pows}}
\end{figure*}

In Figure~\ref{fig:lcs}, the long-term radio, mm, and \gr\ light curves
of \ngc\ are shown. We employed the WWZ and LSP analysis methods to search
for periodicities in the radio, 870-$\mu$m, and \gr\ light curves, and
the results are shown in
Figure~\ref{fig:pows}. The time periods were from
the beginning of the data to 2015 August 01, before the probable C3-clump collision.
No significant signals in the three bands were found. We also conducted the
analysis of the entire light curve data for each of the three datasets, and
the results were the same.

\section{Analysis checks for the QPO signal in NGC~1275}
\label{sec:a3}
\begin{figure*}
\centering
\epsscale{1.0}
	\includegraphics[width=0.43\linewidth]{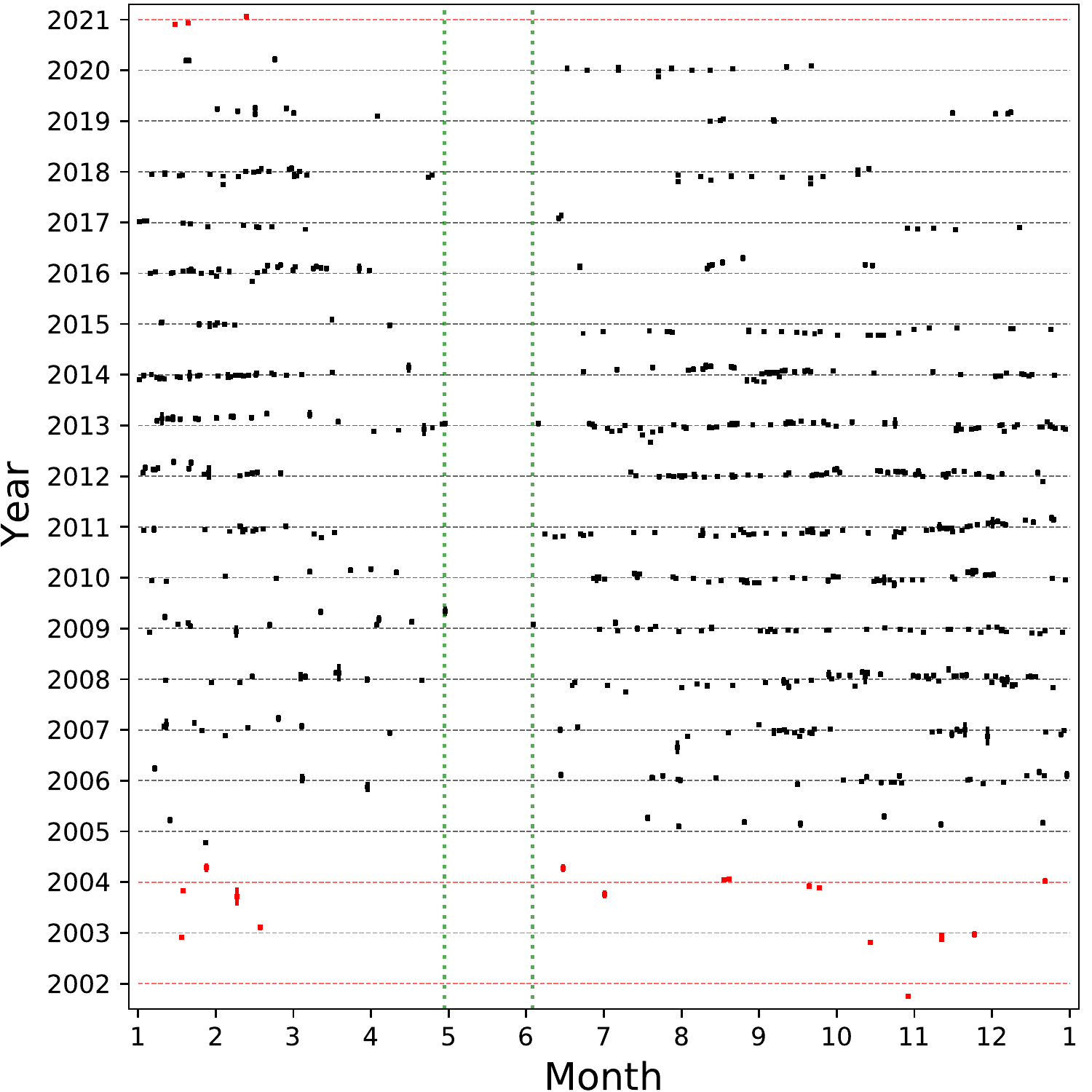}
	\includegraphics[width=0.43\linewidth]{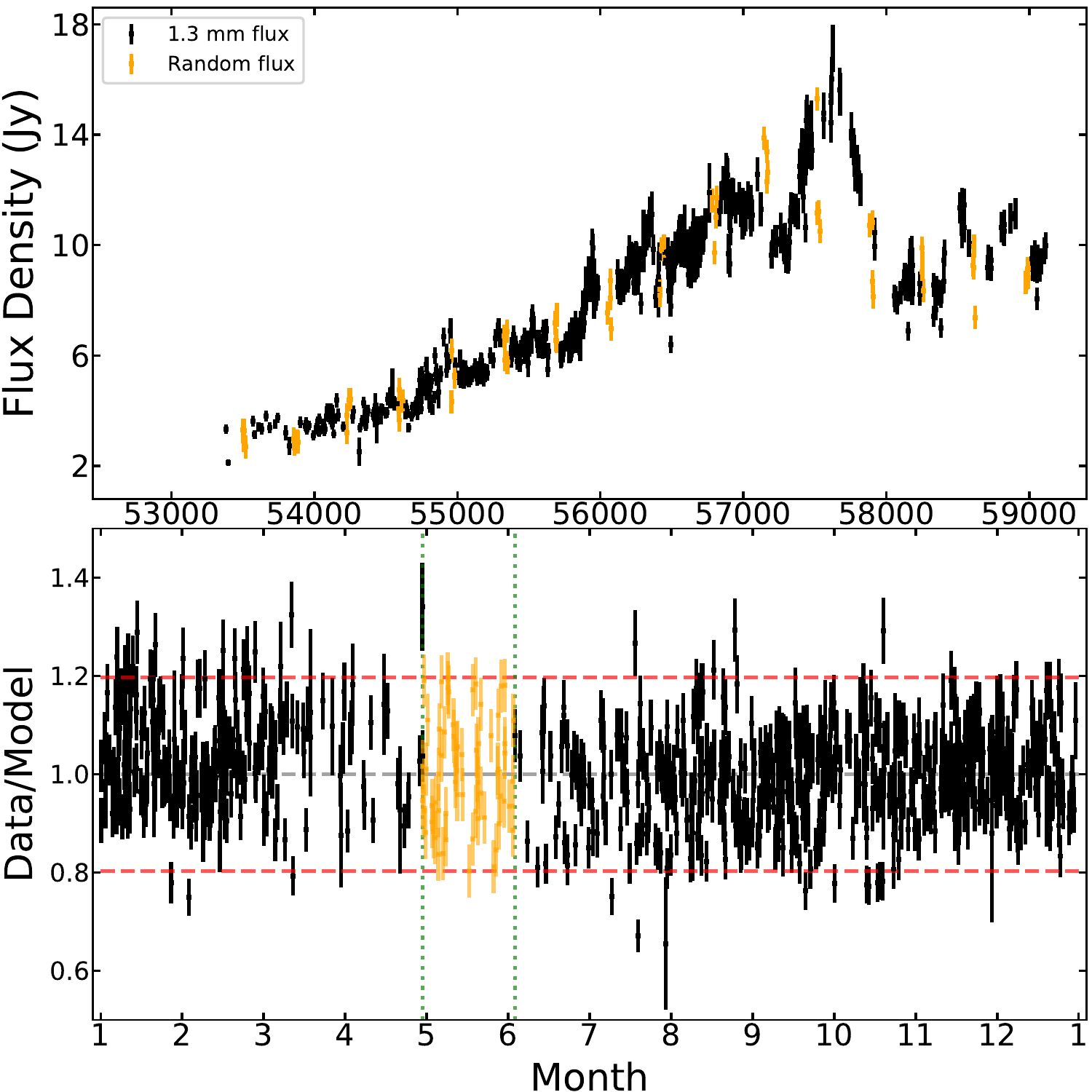}
	\includegraphics[width=0.43\linewidth]{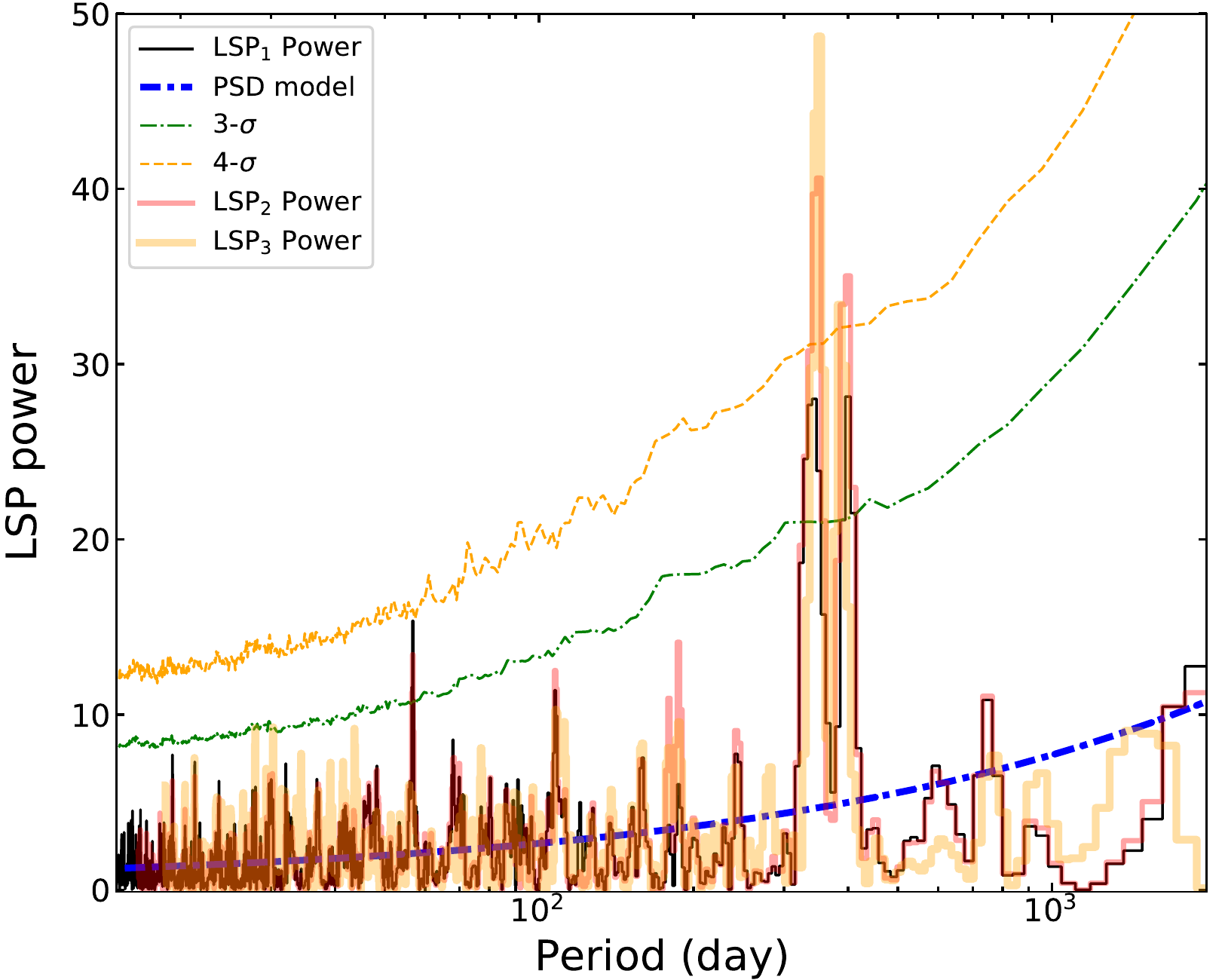}
\caption{{\it Top left}: detrended 1.3-mm light curve of NGC~1275. A gap in May of
	each year (marked by green dotted lines) is clearly seen.
	In addition, there are only a few data points in 2002--2004 and 2021
	(shown in red).
	{\it Top right}: gap-filled light curves for NGC~1275 in 2005--2020.
	For each gap in
	2005-2020, we randomly generated data points based on the mean number
	of data points per month (i.e., setting times in each gap) and 
	$\pm 2\sigma_d$ range (i.e., setting fluxes at those times). The generated data points
	are shown in yellow in the yearly folded light curve (bottom
	panel).
Adding back the values calculated from the detrending functions to the generated
	data points, we obtained the gap-filled 2005--2020 light curve.
	{\it Bottom}: power curves from LSP analyses of the original 1.3-mm
	light curve (yellow), 2005--2020 light curve (orange), 
	gap-filled 2005--2020 light curve (black). The best fit to the black
	power curve is shown as the blue dash-dotted curve, based on which the
	3$\sigma$ (green dashed) and 4$\sigma$ (yellow dotted) significance 
	curves were obtained.
	\label{fig:test}}
\end{figure*}

As shown in Figure~\ref{fig:test}, the 1.3-mm light curve of NGC~1275
has an unavoidable gap in May of each year when the source is within
a 25\,degree radius from the Sun (the solar avoidance zone
for SMA observations), which might cause artifact signals in analyses 
for detecting periodic signals.
In order to investigate whether this observational pattern
could induce the signal we have found, 
we conducted the following analyses as additional checks for the signal. 
Here we include the light curve data after 2015 August 01, as the gaps 
are present throughout the entirity of the data.

\begin{figure*}
\centering
\epsscale{1.0}
	\includegraphics[width=0.43\linewidth]{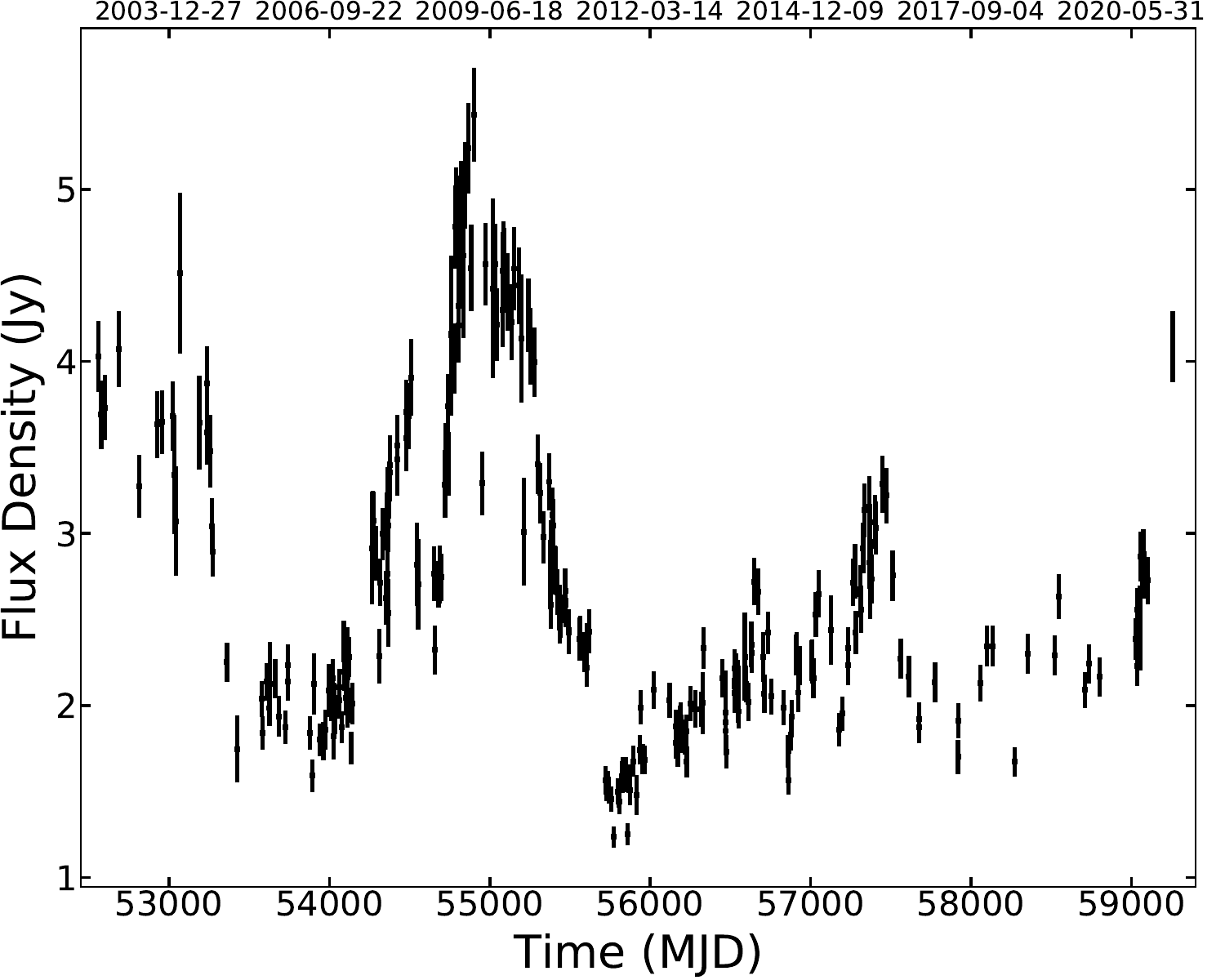}
	\includegraphics[width=0.43\linewidth]{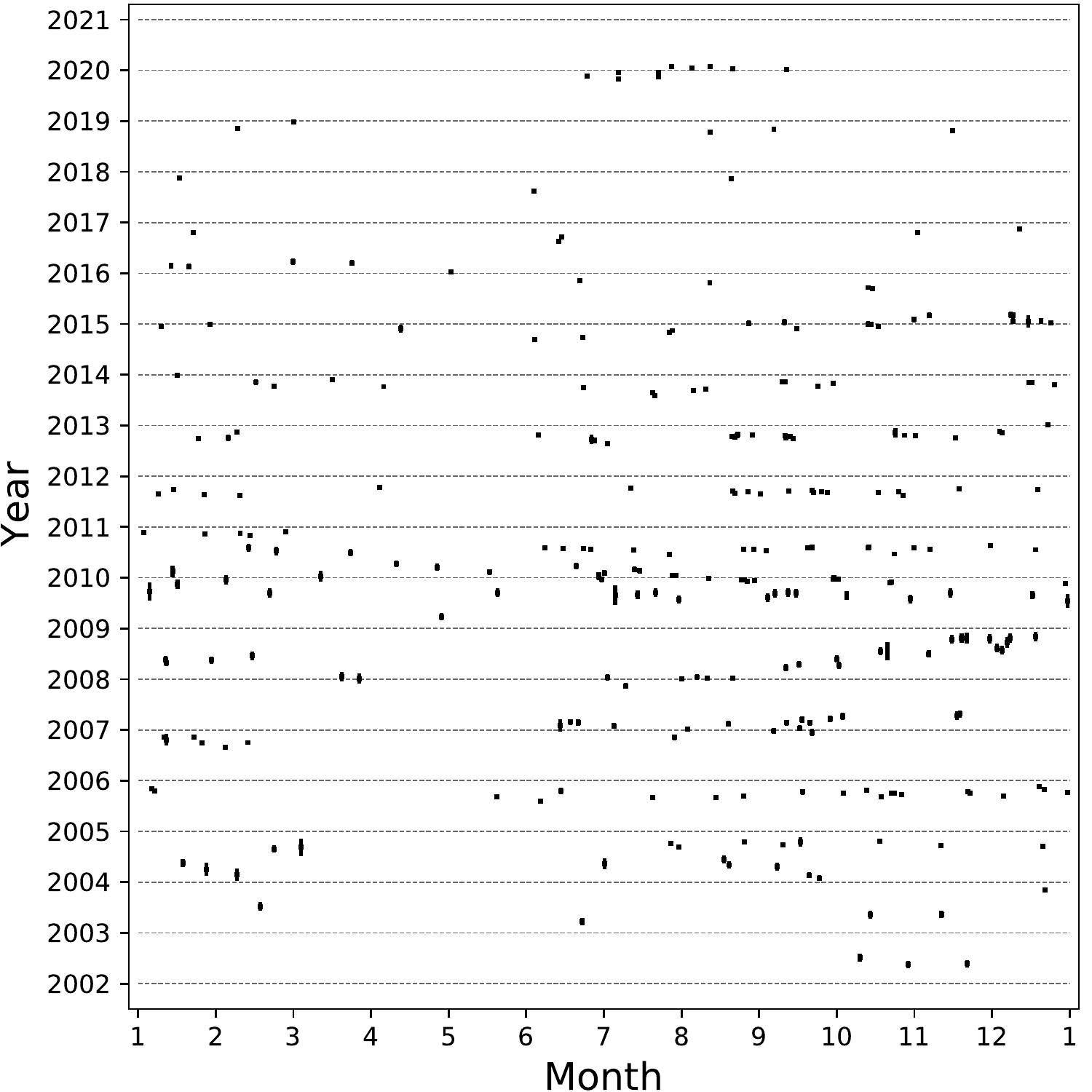}
	\includegraphics[width=0.43\linewidth]{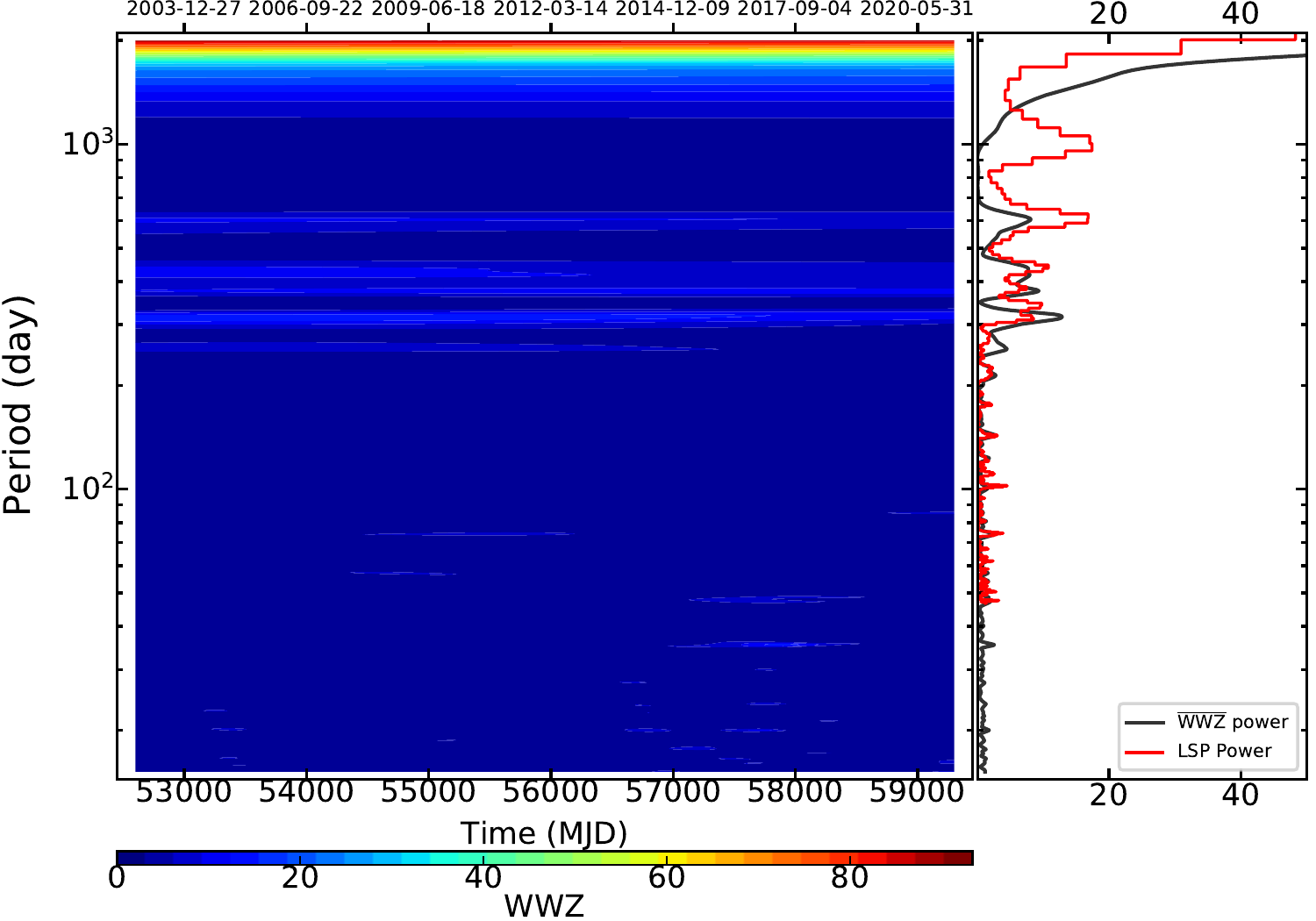}
	\caption{{\it Top figures}: 1.3-mm SMA  light curves of J0359+509, 
	which is close to NGC 1275 in the sky,
	where the left one shows the standard view of the observations 
	and the right one shows the yearly 
	light curve, illustrating a gap in approximately April--May of each year
	(similar to that of NGC 1275).
	{\it Bottom figure}: WWZ power (left panel) and LSP power and 
	summed WWZ power (red and black lines respectively in the right panel)
	obtained for the 1.3-mm light curve.  No signals of QPOs, real or 
	artifactual, are present.
\label{fig:nearby}}
\end{figure*}

\begin{figure*}
\centering
\epsscale{1.0}
	\includegraphics[width=0.43\linewidth]{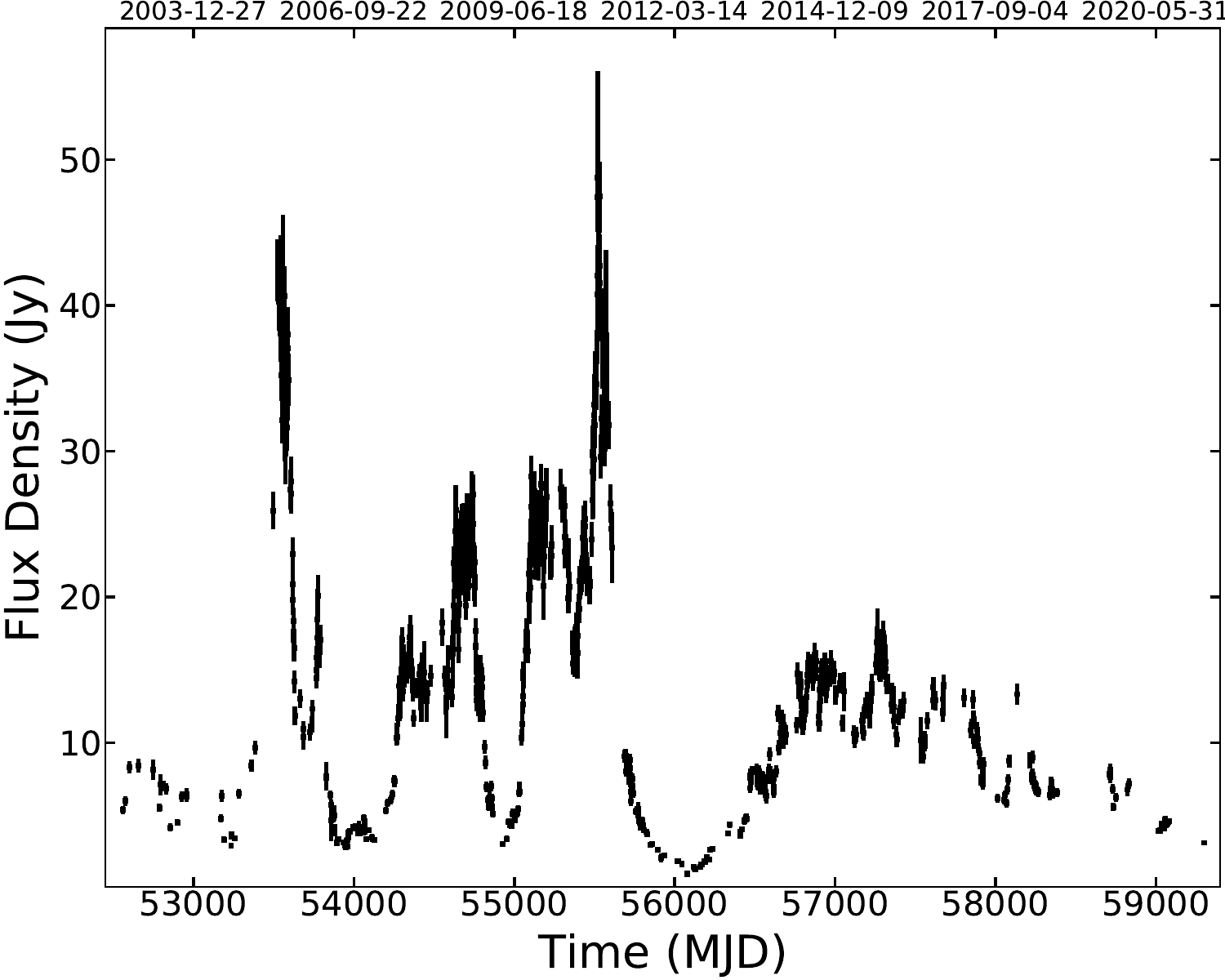}
	\includegraphics[width=0.43\linewidth]{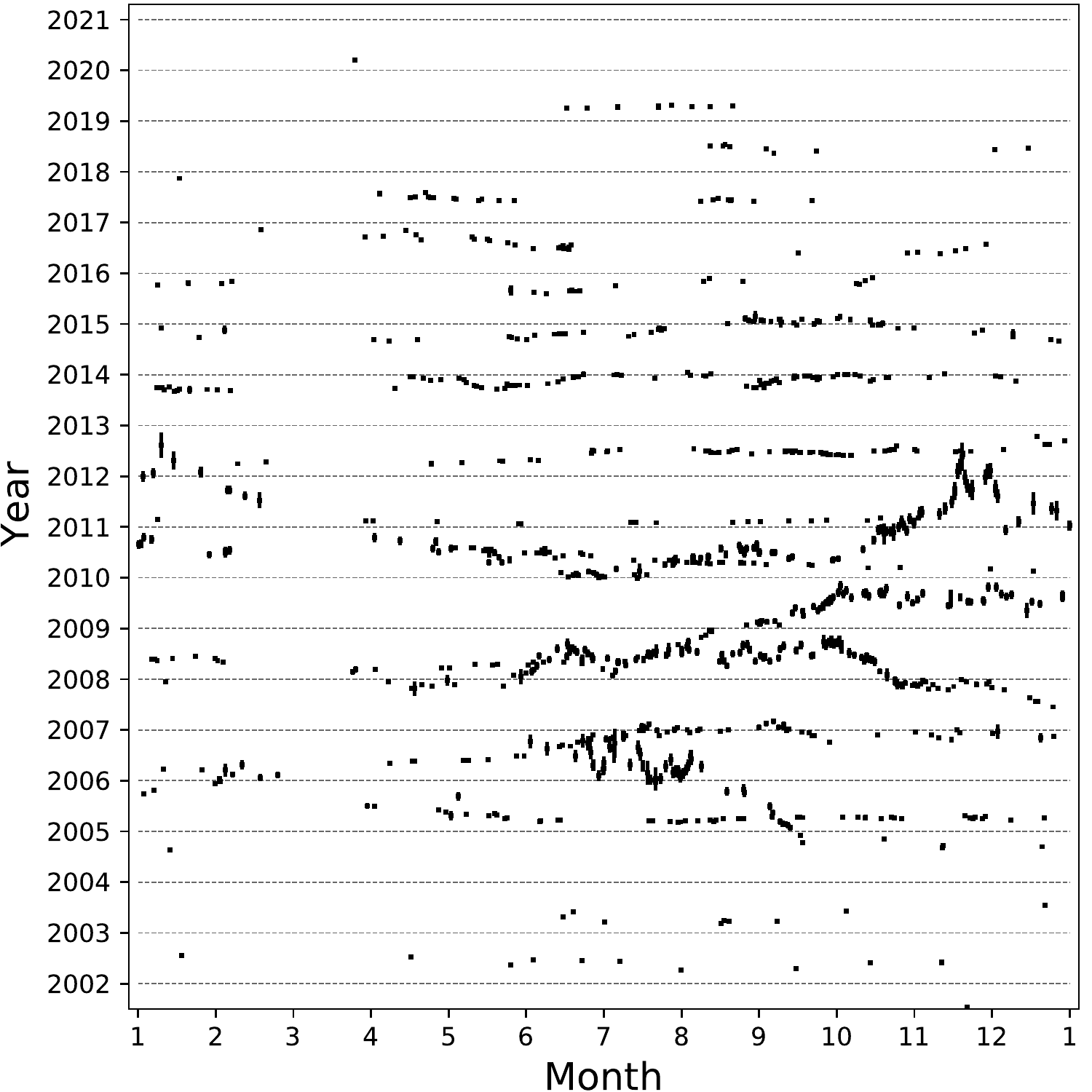}
	\includegraphics[width=0.43\linewidth]{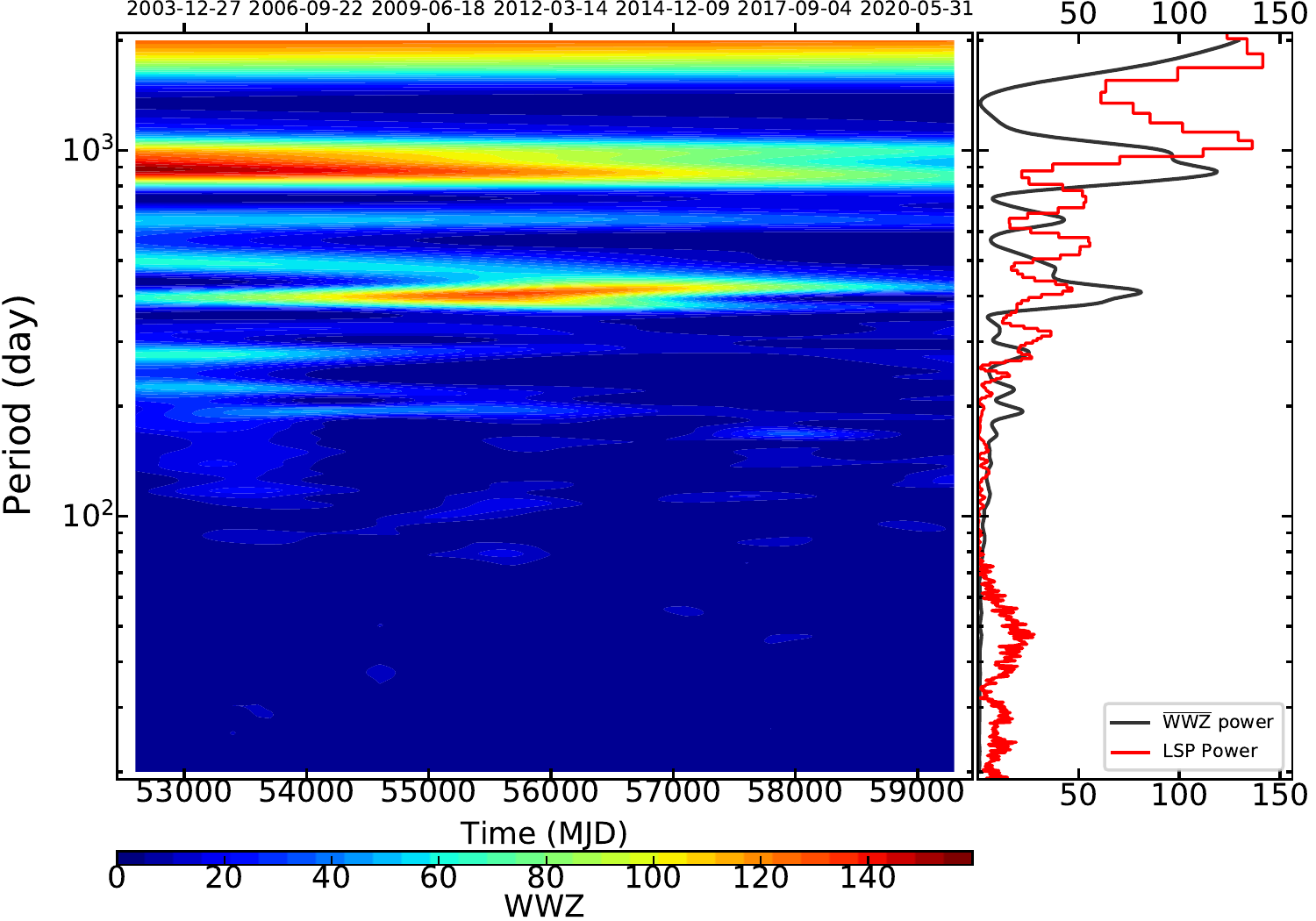}
	\caption{{\it Top figures}: 1.3-mm SMA light curves of J2253+161, 
	a very well measured source, where
	the left one shows the standard view of the observations and the right 
	one shows the yearly light curve.
	There is a gap in March of each year.
	{\it Bottom figure}: WWZ power (left panel) and LSP power and 
	summed WWZ power (red and black lines respectively in the right panel)
	for the 1.3-mm light curve. No signals could be identified with a
	sufficiently high significance and, critically, none are present 
	at the periods seen in the light curve of NGC 1275.
\label{fig:freq}}
\end{figure*}

Firstly we conducted the same analysis to the 1.3-mm flux density light curves 
of J0359+509
(R.~A.=$03^h59^m29\fs7472$, Decl.=$50^{\circ}57^{'}50\farcs161$)
and J2253+161
(R.~A.=$22^h53^m57\fs7479$, Decl.=$16^{\circ}08^{'}53\farcs560$),
obtained
with the SMA in the same monitoring program. The former is a source close 
to our target in the sky 
and the latter was one of the most frequently observed with the SMA, 
and so provides a
particularly good test for production of any artifactual signal.
For J0359+509, a paucity of data points in approximately April--May of 
each year is present 
probably for the same reason that NGC 1275 could not be observed then, but
in total there are 287 data points in its light curve (Figure~\ref{fig:nearby}).
For J2253+161, there is an obvious gap in March of each year but its light curve
has 919 data points (Figure~\ref{fig:freq}). From the WWZ and LSP analyses, 
no significant signals that were
 similar to those appearing in the power spectra of NGC~1275 were seen.

We further investigated how the gap might affect the QPO signal in NGC~1275.
Examining the yearly light curve (of the detrended one), we noted that
there were only a few data points in 2002--2004 and 2021. Ignoring these
sparse
data points, the LSP power peak of $P_l$ would be lowered from $\sim 50$ to
$\sim 40$, while the power peak of $P_h$ would be kept nearly the same
(Figure~\ref{fig:test}). We estimated the mean number
of the data points per month ($\simeq$4) based on the 2005--2020 light curve,
and obtained its standard deviation ($\sigma_d$). We then 
generated randomized pseudo-data points to fill in the May gaps in this 
light curve.
The times of these data points were randomly chosen in each May and their fluxes
were randomly chosen within the $\pm 2\sigma_d$ range (the average flux 
uncertainty was given to each generated flux). The generated data points
(64 in total) are shown in the top right panel of Figure~\ref{fig:test}. 
Using the two
functions for detrending the long-term variations,
we brought back the generated data points to fit into the original light curve 
for
2005--2020.  We then conducted a LSP analysis of this gap-filled light curve 
and the resulting power spectrum density (PSD) is shown in the bottom panel of 
Figure~\ref{fig:test}. Both the $P_l$ and $P_h$ periodicity peaks were lowered 
to $\sim 30$. We modeled the PSD and then generated 5$\times 10^5$ light curves
based on the model fit. The 3$\sigma$ and 4$\sigma$ significance curves were
obtained from the simulated light curves. The two power peaks remain 
significant at a $>3\sigma$ level.

We understand the above results as follows.
As the May gaps are the major pattern seen in the observed light curve, we 
tested to remove the pattern by filling the generated data points. 
The LSP 
analysis results still show that the two-periodicities are significant,
although the significances are lowered, particularly the $P_l$ one. Since
we generated approximately 9\% of the observed data points, for which
their fluxes were randomly chosen from the $\pm 2\sigma_d$ range around 
the smoothly fitted underlying trends, we 
effectively added noise to any putative QPO signal present in the light curve. 
Thus the 
significances naturally became lower than those in the observed light curve,
but the signal remains strong.
We note that a weak peak in the LSP PDS at $\sim 57$\,d is slightly increased 
instead,
now reaching nearly the 4$\sigma$ significance curve (Figure~\ref{fig:test}).
The reason for this change is not clear, but in any case we mostly modified 
the low-frequency
structures of the PSD by adding the data points and the high-frequency 
structures might not be significantly affected.

From these analyses, by searching for periodic signals from a nearby source
and a most frequently observed source whose light curve has a similar gap
pattern, and by filling the gaps in the light curve 
of NGC~1275 with random data points, we conclude that the two periodicities we have found are real
signals, not artifacts induced by the constrained observation patterns.

\end{document}